\documentclass[conference]{IEEEtran}

\newcommand{\arevised}[1]{\textcolor{blue}{#1}}

\newcommand{\drevised}[1]{\sout{#1}}

\renewcommand{\arevised}[1]{#1}

\renewcommand{\drevised}[1]{}

\newcommand{\subject}[1]{\vspace{3pt}\noindent\textbf{#1}}
\newcommand{\subsubject}[1]{\vspace{3pt}\noindent\textit{#1}}

\newcommand\malurl[1]{\href{notalink}{{\nolinkurl{#1}}}}
\newcounter{finding}
\newcommand{\finding}[1]{
\vspace{3pt}
\noindent
\framebox{
\begin{minipage}[b]{3.3in}
\noindent \textbf{Finding \Roman{finding}}: \textit{#1}
\stepcounter{finding}
\end{minipage}
}
\vspace{-5pt}
}
\renewcommand{\finding}[1]{}

\newcommand{\ignore}[1]{}
\usepackage{CJKutf8}
\usepackage{amsmath}
\usepackage{graphicx}
\usepackage[tight,footnotesize]{subfigure}
\usepackage{multirow}
\usepackage{fancyvrb}
\usepackage[flushleft, online]{threeparttable}
\usepackage{xurl}
\usepackage{array}
\usepackage{algorithmic}
\usepackage{booktabs}
\usepackage{xcolor}
\usepackage{float}
\usepackage{ulem}
\ifdefined\ccs
\else
    \usepackage{cite}
\fi
\usepackage{enumitem}
\usepackage[colorlinks,
            linkcolor=blue,
            anchorcolor=blue,
            citecolor=blue,
            urlcolor=blue
            ]{hyperref}

\begin{document}

\title{Reflected Search Poisoning for Illicit Promotion}

\makeatletter
\newcommand{\linebreakand}{%
  \end{@IEEEauthorhalign}
  \hfill\mbox{}\par
  \mbox{}\hfill\begin{@IEEEauthorhalign}
}
\makeatother



%
\author{\IEEEauthorblockN{Sangyi Wu,
Jialong Xue,
Shaoxuan Zhou and
Xianghang Mi }
\IEEEauthorblockA{School of Computer Science and Technology\\
University of Science and Technology of China\\ 
\url{https://chasesecurity.github.io/Reflected-Search-Poisoning-for-Illicit-Promotion}}
}

\maketitle

\begin{abstract}
As an emerging black hat search engine optimization (SEO) technique, reflected search poisoning (RSP) allows a miscreant to free-ride the reputation of high-ranking websites, poisoning search engines with illicit promotion texts (IPTs) in an efficient and stealthy manner, while avoiding the burden of continuous website compromise as required by traditional promotion infections. However, little is known about the security implications of RSP, e.g., what illicit promotion campaigns are being distributed by RSP, and to what extent regular search users can be exposed to illicit promotion texts distributed by RSP. In this study, we conduct the first security study on RSP-based illicit promotion, which is made possible through an end-to-end methodology for capturing, analyzing, and infiltrating IPTs.

As a result, IPTs distributed via RSP are found to be large-scale, continuously growing, and diverse in both illicit categories and natural languages. Particularly, we have identified over 11 million distinct IPTs belonging to 14 different illicit categories, with typical examples including drug trading, data theft, counterfeit goods, and hacking services. Also, the underlying RSP cases have abused tens of thousands of high-ranking websites, as well as extensively poisoning all four popular search engines we studied, especially Google Search and Bing. Furthermore, it is observed that benign search users are being exposed to IPTs at a concerning extent. To facilitate interaction with potential customers (victim search users), miscreants tend to embed various types of contacts in IPTs, especially instant messaging accounts. Further infiltration of these IPT contacts reveals that the underlying illicit campaigns are operated on a large scale. All these findings highlight the negative security implications of IPTs and RSPs, and thus call for more efforts to mitigate RSP-driven illicit promotion.

\end{abstract}

\begin{CJK*}{UTF8}{gbsn}
    \section{Introduction}
\label{sec:intro}
\newcommand{\rsp}{RSP}
\newcommand{\rspL}{reflected search poisoning}
\newcommand{\cp}{CP}
\newcommand{\cpL}{cybercrime promotion}
\newcommand{\detector}{mole}
Search engine optimization (SEO) techniques are widely adopted by website administrators to enhance the visibility of their websites to search engine users. Typical legitimate SEO techniques include on-page optimization (e.g., link building and forum posting) and off-page optimization (e.g., enhancing robots.txt or sitemap), which can guide search engines in matching search queries with webpages that are most relevant. However, as the dark side of SEO,  black hat SEO activities emerged decades ago and have since been actively conducted by miscreants to mislead search engines, gain unfairly high page rankings, and expose search engine users to unsolicited or even harmful webpages. Typical black hat SEO techniques include forum spamming~\cite{niu2007quantitative}, link farms~\cite{wu2005identifying} and spider pooling~\cite{du2016ever}. A black hat SEO activity is considered as search engine poisoning or simply \textit{search poisoning} when it is used to promote illicit and harmful content, and most search engine poisoning attacks are conducted through promotional infections which involve the injection of mostly malicious webpages into a compromised website. Given the prominent threat of promotional infections, significant efforts~\cite{john2011deseo, liao2016seeking} have been invested into profiling and detecting them. Also, since promotional infections can be quickly recovered once detected~\cite{leontiadis2014nearly}, attackers have to keep compromising new websites so as to maintain the magnitude of their search poisoning campaigns. Besides, most high-ranking websites tend to be well-protected and are thus immune to such promotional infections. All these factors motivate miscreants to identify alternative search poisoning techniques that are more sustainable and more efficient. 

In recent years, a novel and understudied search poisoning technique has emerged, gaining noticeable adoption among miscreants, which we name as Reflected Search Poisoning (RSP). At its core, RSP enables miscreants to inject harmful content into the pages of benign and popular websites, deviating from the original intent of the webpage design, free riding reputation of the website under abuse,  and thereby misleading search engines into indexing the injected harmful content with a high page rank. This technique capitalizes on the dynamic nature of most websites, wherein they \textit{reflect} one or more URL text parameters into the rendered webpage content. We denote such URL patterns as \textit{URL Reflection Schemes} (URSes), with the corresponding URL parameters referred to as \textit{reflection parameters}. \arevised{Depending on the position of the parameters being reflected in the webpage, seven reflection methods have been identified through our manual study. The most common reflection targets include the webpage title, input box values, and plain text for display purposes only. Reflection also extends to webpage metadata, JavaScript code variables, anchor links, and customized data attributes. }
For instance, \textit{https://www.youtube.com/results}  allows YouTube users to specify search terms using the URL parameter \textit{search\_query}. When \textit{https://www.youtube.com/results?-search\_query=reflection-text} is specified, the URL parameter value \textit{reflection-text} will be reflected as the title of the resulting webpage, even if no search results are found. 

Given the URL reflection schemes (URSes) of high-rank websites, a typical RSP attack starts by assigning an URS with illicit promotion texts, which leads to RSP URLs generated. Then, these RSP URLs will be distributed across the Internet using traditional blackhat SEO techniques such as forum spamming\cite{niu2007quantitative} and spider pools\cite{du2016ever}, and finally get indexed by search engines. 
Since an RSP attack does not involve the compromise of victim websites, it leaves no files on the victim web servers except for the visiting logs from search engines, and is thus more stealthy and less costly. Also, since URL reflection schemes are widely available on popular websites, the RSP attack allows miscreants to free-ride the reputation of popular websites and gain high page rankings for the poisoned search entries.
As illustrated in Fig.~\ref{fig:search_CPT_introduction}, when querying Google Search with the Chinese keyword "美国文凭" (USA diploma) on Sep 26, 2023, all of the top 20 search results were found to be cases of reflected search poisoning (RSPs) promoting fake certificate services. Also, these illicit promotion texts (IPTs) were observed on multiple high-ranking websites, such as \textit{yahoo.com} and \textit{azurefd.net}.
\begin{figure}
\centering
\includegraphics[width=.9\columnwidth]{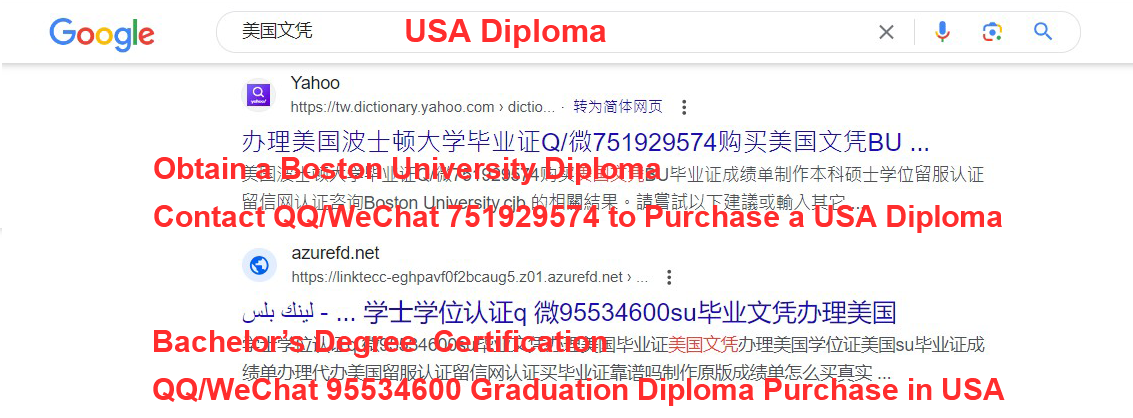}
\caption{Reflected search poisoning cases promoting fake certificate services got returned when querying Google Search with the Chinese keyword "美国文凭" (USA diploma) on Sep 26, 2023. 
}
\label{fig:search_CPT_introduction}
\end{figure}

In this study, we aim to answer the following research questions. First of all, \textit{what illicit promotion activities are being distributed through RSP?} Also, \textit{to what extent have RSP attacks poisoned search engines and abused popular websites?} Besides, \textit{how likely can regular search users be exposed to RSP-based illicit promotion?} Furthermore, given victims exposed to IPTs, \textit{how would they further interact with the underlying operators of IPTs, i.e., sellers of illegal services or goods?}





To answer these questions, multiple technical challenges must be tackled. Firstly, it is necessary to effectively hunt IPTs and the underlying RSP cases with high precision and good coverage.
Particularly, the hunter should be able to distinguish RSPs from benign search entries especially legitimate URL reflection cases. Furthermore, no tools are available to automatically classify large-scale IPTs into well-defined categories, which impedes an in-depth understanding of illicit promotion activities.
Furthermore, most IPTs have contact entities embedded as the next hops for the communication with victims, e.g., instant messaging accounts and websites. However, it turns out to be challenging to automatically extract such contacts due to various evasion techniques deployed by the operators, e.g., using homoglyphs and inserting meaningless characters. 

We address these challenges through the design and implementation of three novel tools. The first is \textit{the IPT hunter} which is designed to query search engines with IPT-relevant keywords and URL reflection schemes, and automatically distinguish IPTs from benign search entries by means of a binary IPT classifier.
The second tool is \textit{the IPT analyzer}, which can classify IPTs into pre-defined categories and extract out embedded contacts if any. Then, the third tool, namely, \textit{the IPT infiltrator},  is designed to carry out in-depth infiltration for the extracted IPT contacts so as to better profile the underlying operators of illicit promotion campaigns. The key findings of this study are highlighted as below.

First of all, \textbf{
RSP-based IPTs are large-scale and multilingual, involve diverse categories of illicit goods and services, and have extensively poisoned all four search engines under this study.} In total, our IPT hunter has discovered around 11.9 million distinct IPTs and 13.3 million RSP cases, which however can only be considered as a lower-bound estimate for RSP-based IPTs since our IPT hunter is subject to strict rate limits of the search engines and was deployed for less than three months in total. 
Then, these RSP cases were indexed by at least one of the four popular search engines under our study: Google, Bing, Baidu, and Sogou.
Besides, the observed IPTs belong to a diverse set of over 14 categories of illicit goods and services which range from well-known ones (e.g., illegal drug sales, counterfeit goods, and gambling) to emerging ones (data theft and hacking services).  Also, IPTs in each category have diverse goods or services observed.  For instance, in our small-scale sampling alone, fraudulent accounts of 20 different popular online services are on sale through IPTs while 13 distinct drugs were observed in drug IPTs.

Also, \textbf{high-rank websites have been heavily abused in RSP-based illicit promotion, while regular search engine users can be exposed extensively to various IPTs.} 
Among 60,638 benign websites (apex domains) abused in RSP attacks, 20,330 rank in the top 1M most popular websites, 2,113 belong to the top 10K, 854 are renowned education institutions, and 1,144 are websites of government agencies. 
We also comprehensively profiled the extent to which regular search engine users can be exposed to RSPs and IPTs. And it turns out that at least three categories of search keywords are likely to expose search users to IPTs:  location names (e.g., city names), benign long-tail keywords, and terms of various illicit goods or services. Particularly, when querying Google Search with one of the most popular 3,368 city names in China, 46\% search queries will have one or more IPTs ranked in the top 10 result entries.

In addition, \textbf{campaigns and operators underpinning IPTs turn out to be large-scale and tend to redirect potential customers to instant messaging platforms for further communication and trading.}
83.62\% IPTs have instant messaging contacts embedded as the next hop to further interact with victims (potential customers), while the rest redirect victims to their websites. 
Leveraging the IPT analyzer, we have extracted 48,114 IPT contacts in total which consist of 16,335 websites, 5,890 Telegram accounts, 23,632 WeChat accounts, 1,552 QQ accounts, and 705 telephone numbers.
And our infiltration on these contacts has distilled a series of interesting observations, e.g., novel evasion techniques employed by websites and instant messaging accounts, the distribution of mobile apps for child pornography and illegal gambling, and the extensive communication traces observed on the Telegram platform for illicit goods and services. 

We have responsibly disclosed our findings to the four search engines. Among them, Bing has responded to our disclosure and acknowledged the reported issues, while we have yet to receive any concrete response from the other three search engines. Also, Bing appears to have taken effective mitigation actions since our disclosure, as a significant drop in RSP-based IPTs is observed for Bing when comparing our first-phase IPT hunting in November 2022 with the second-phase hunting in November 2023.


Our contributions can be summarized as below.

\vspace{2pt}\noindent$\bullet$ To our best knowledge, this is the first extensive study on illicit promotion through reflected search poisoning. Our study has observed RSP-based IPTs at a large scale, upon which a set of novel observations have been further distilled. 

\vspace{2pt}\noindent$\bullet$ Three novel tools have been designed and implemented, which allow any party that is independent from search engines to capture, analyze, and infiltrate reflected search poisoning for illicit promotion. 
\section{Background and Related Works}
\label{sec:related}

\subject{The underground economy and illicit promotion}. Over the past two decades, a continual research effort has been invested in the underground economy, i.e., the buying and selling of illicit goods or services. Among these works, a large portion is dedicated to specific categories of illicit services and goods, e.g., illicit or even illegal drug trading~\cite{kanich2011show, leontiadis2011measuring, mccoy2012pharmaleaks, aldridge2017delivery, li2021demystifying}, 
illegal online gambling~\cite{yang2019casino, gao2021demystifying}, malware distribution~\cite{caballero2011measuring, kotzias2016measuring, thomas2016investigating, kotzias2021did}, among others.

Another line of studies focuses on holistic infrastructures that are used in the promotion of illicit goods and services, i.e., illicit promotion. 
Particularly, Grier et al.~\cite{grier2010spam} revealed in 2010, the promotion of spam URLs on Twitter and found that 8\% URLs posted on Twitter pointed to malicious websites (e.g., malware, and scams). Furthermore, following the emergence of the Tor anonymity network, Christin et al.~\cite{christin2013traveling} conducted a measurement study between 2011 and 2012 on the Silk Road, an anonymous marketplace running as an onion service. As revealed by this study, the Silk Road marketplace was mostly used to trade illegal drugs and most items on sale are available for less than three weeks. 
As a follow-up, Soska et al.~\cite{soska2015measuring} moved to profile the longitudinal evolution of anonymous marketplaces between 2013 and 2015. Still, most sales were found to be drug-related.

Besides, popular online services have been abused by miscreants for illicit promotion so as to reach regular online users. One typical channel is promotional infections~\cite{john2011deseo, leontiadis2014nearly, liao2016seeking} wherein the attacker compromises a legitimate website, injects promotional and harmful webpages, induces the search engines to index these webpages with high page rankings, and ultimately exposes benign search users to their malicious campaigns. And such promotional infections have been used by miscreants to promote illicit activities of diverse categories, e.g., online gambling~\cite{liao2016seeking}, unlicensed pharmacies~\cite{leontiadis2014nearly}, etc. Also, as revealed by \cite{leontiadis2014nearly}, the median time to recover from such promotional infections is around 15 days, which indicates miscreants may have to continuously compromise new vulnerable websites so as to maintain or increase the visibility of their promotion activities. 
Furthermore, John et al.~\cite{john2011deseo} proposed a methodology to detect if a search result is poisoned or not by looking at its URL parameters instead of visiting the URL. Besides, Liao et al.~\cite{liao2016seeking} utilized the semantic inconsistency between the injected promotional content and the legitimate context of the infected website to determine if a webpage is a promotional infection or not.

Besides, some research efforts~\cite{zhao2016chinese, yang2017learn, yuan2018reading, zhu2021self} have focused on identifying and understanding jargon words used in illicit promotion. Particularly, Yang et al.~\cite{yang2017learn} explored how to identify jargon words from keywords promoted in blackhat SEO campaigns,  while Yuan et al.~\cite{yuan2018reading} compared the semantic discrepancy of the
same word between contexts of illicit promotion and benign usage scenarios to identify
jargon words. 

\subject{Search engine poisoning}. 
Search engine poisoning (i.e., search poisoning) encompasses blackhat SEO activities that are conducted by miscreants with the goal of poisoning the search results with \textit{harmful content}~\cite{wang2013juice}. When carrying out search poisoning attacks, miscreants usually deploy various
blackhat SEO techniques, e.g., forum spamming~\cite{niu2007quantitative}, cloaking~\cite{wang2011cloak, invernizzi2016cloak}, spider pooling~\cite{du2016ever}, content spinning~\cite{zhang2014dspin}, suggestion manipulation~\cite{wang2018game}, and promotional infections~\cite{john2011deseo, liao2016seeking}. Also, a series of research works have been devoted to detecting search poisoning through promotional infections~\cite{john2011deseo, lu2011surf, liao2016seeking, yang2021scalable}. Particularly, John et al.~\cite{john2011deseo} focused on clustering compromised webpages into their underlying poisoning campaigns, and Lu et al.~\cite{lu2011surf} proposed SURF to capture poisoned search results after observing that the redirection chain of visiting a poisoned search result has more distinct features compared with visiting a benign one. 

Compared with these previous works on illicit promotion or search poisoning, this study focuses on an emerging search poisoning technique (reflected search poisoning) that is mainly used for illicit promotion towards regular online users. Particularly, it quantitatively reveals, for the first time, how this new search poisoning technique is being extensively used to conduct stealthy and effective illicit promotion for goods and services belonging to diverse categories. 



\section{Methodology}
\label{sec:method}

\begin{figure}
    \centering
    \includegraphics[width=.98\columnwidth, height=.18\textwidth]{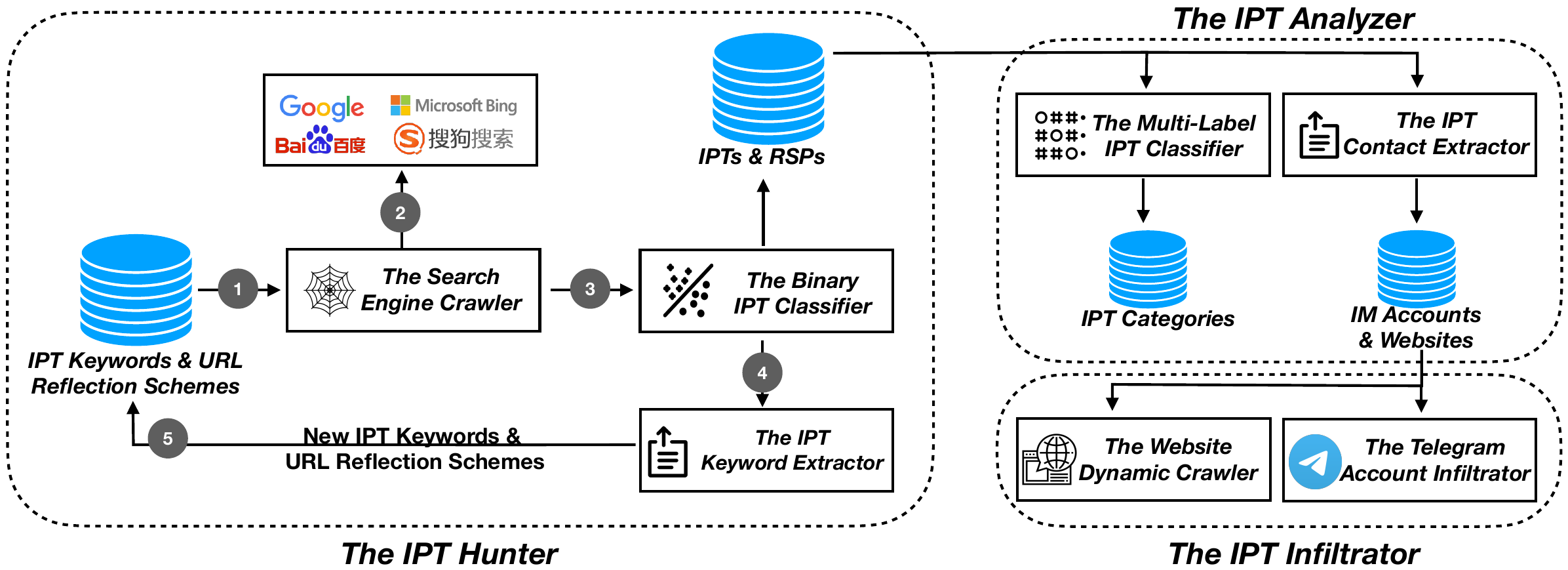}
    \caption{The methodology to detect and understand RSP-based IPTs.}
    \label{fig:methodology}
\end{figure}

In this section, we introduce our methodology to discover and profile illicit promotion through reflected search poisoning. As shown in Fig.~\ref{fig:methodology}, our methodology consists of three major components. One is to discover IPTs and RSPs through the IPT hunter ( \S\ref{subsec:method_hunter}). Then, given IPTs, an analyzer (\S\ref{subsec:method_cpt_analyzer}) is further applied to profile IPT categories and extract contact identifiers as embedded in IPTs, through which, a large volume of contacts  including instant messaging accounts and websites have been discovered. To reveal what these contacts will redirect a victim to, an IPT infiltrator (\S\ref{subsec:method_cpt_infiltrator}) is designed to automatically visit and profile  websites and Telegram accounts that are promoted in IPTs.



\subsection{The IPT Hunter}
\label{subsec:method_hunter}

Our IPT hunter is built upon two observations. One is that the same IPT can be distributed in RSP cases across different URL reflection schemes. On the other hand, the same URL reflection scheme can be abused to promote IPTs of different campaigns and different categories of illicit goods and services. Upon these observations, our IPT hunter is designed to identify RSP-based IPTs through three steps. 
The first step is to query search engines with well-crafted text keywords and URL reflection schemes, leading to the discovery of potential RSP cases. Then, to distinguish RSP cases from benign search entries (e.g., legitimate cases of URL reflection schemes), the second step applies a machine learning classifier to automatically decide whether the URL of a given search result entry contains an IPT or not. Once benign cases get filtered out, the third step is to extract new search keywords and URL reflection schemes before feeding them into the first search step, so as to snowball the discovery process.

\subject{The search engine crawler.}
As discussed above, the first step of our IPT hunter is to query search engines, for which, a crawler is built up along with two search strategies implemented. One is to query a search engine with keywords extracted from known IPTs, which leads to the discovery of RSP cases serving similar IPTs but across different URL reflection schemes. Besides, 
upon the observation that the same URL reflection scheme can be abused by different IPT campaigns, 
the other strategy queries search engines for result entries that match a given URL reflection scheme, which gives IPTs that are hosted on the same URL reflection scheme but belong to different campaigns and categories. Therefore, iterative querying search engines with these two strategies allows us to effectively extend to new RSPs/IPTs as well as new URL reflection schemes that are abused in RSPs.

To implement the first keyword-based search strategy, one obstacle resides in that most IPTs are lengthy with a median length of 46 characters and querying with such a lengthy IPT typically confuses the search engine and has a low hit rate for new RSPs.
Instead, as detailed below, an IPT keyword generator is built up to split an IPT and locate its most critical component which will be used as the search keyword (i.e., the IPT keyword). 
Regarding the second URL-based search strategy, 
it is only enabled for Google Search since the other three search engines don't support filtering search results with an URL scheme. When searching Google, the \textit{site} filter is set up with a given URL reflection scheme (URS).
 Besides, since Google Search limits the maximum number of search results returned for each query (between 300 and 400), when a URS has many benign reflection cases, searching with it can likely have benign reflection entries dominating the search results. Therefore, for each URS, in addition to the vanilla search query of \textit{site:URS}, alternative search queries are designed to further surface RSP cases. All these alternative queries feature the combination of \textit{site:URS} and a keyword from the following well-crafted keyword set \{url, tg, telegram, 微, 薇, 扣, qq, 飞机\}. Such keywords are crafted based on the observation that RSP cases tend to have contacts or websites embedded, and these contact-relevant keywords (e.g., \textit{tg} for Telegram, and \textit{qq} for QQ) make them distinguishable from benign reflection cases. Our preliminary experiments have demonstrated the effectiveness of such search strategies in terms of discovering IPTs even when there are many benign reflection cases for a given URL reflection scheme. 

The result search entries can still be a mix of RSPs, legitimate URL reflection cases, as well as other benign search entries. Therefore, a filtering step is further applied to excluding search results that are not URL reflection cases. This is achieved by a simple but accurate filtering rule, wherein a search result entry is considered as URL reflection only when the values of one or more URL parameters are rendered as part of the webpage elements. Given URL reflection cases identified, the next step is to distinguish RSPs from benign URL reflections. Since RSPs are mainly used to distribute IPTs, we achieve this goal through a binary IPT classifier.

\subject{The binary IPT classifier.} To build up a binary IPT classifier, the first step is to create a representative ground truth dataset.
To achieve this, we utilized aforementioned search engine crawler and searched Google Search with seeding IPT keywords and URL reflection schemes as learned from manual study. Then, the texts of the resulting URL reflection cases were manually extracted and then labeled. When labeling a text, a text entry is considered as an IPT only when it satisfies all the following criteria: 1) It promotes goods or services, especially illicit ones; 2) Its semantics deviates significantly from the context of the parent webpage and website; 3) Its semantics and syntax deviates significantly from the intended usage scenario for the respective URL reflection scheme. 
Also, when iteratively extending the groundtruth, we prefer new IPTs and non-IPTs that are under-represented in the current groundtruth in terms of IPT categories and natural languages.
As the result, a groundtruth dataset was composed with 2,299 IPTs as positive samples and 1,468 benign URL reflection cases as negative samples.
Each sample, regardless of positive or negative, is a sequence of Unicode text characters, as extracted from  the respective URL reflection case. Also, this ground truth dataset is diverse and representative, as the positive samples cover all the categories of goods and services listed in Table~\ref{tab:cybercrime_categories} and are in 10 different natural languages.

\begin{figure}
    \centering
    \subfigure[An IPT promoting illegal pharmaceuticals. \label{fig:cpt_example_1}]{
        \includegraphics[width=.45\columnwidth]{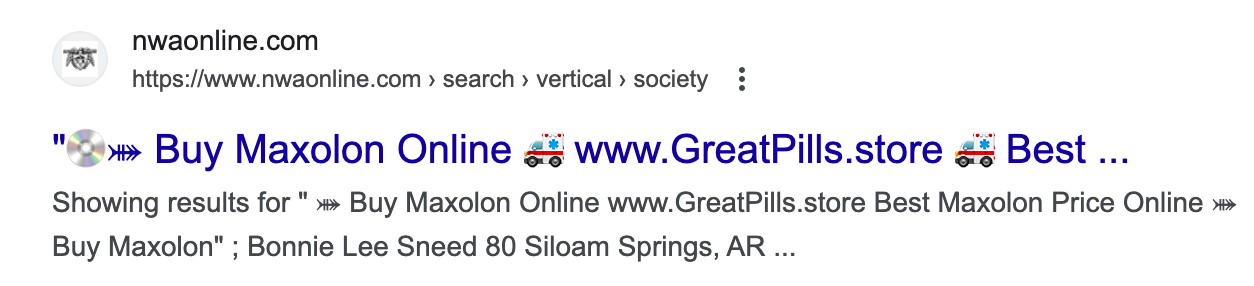}
    }
    \hfill
     \subfigure[An IPT promoting cannabinoid.\label{fig:cpt_example_2}]{
        \includegraphics[width=.45\columnwidth]{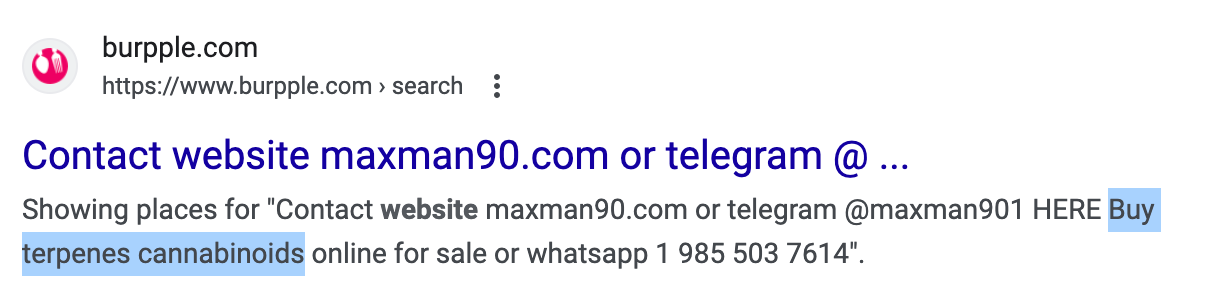}
    }
    \subfigure[An IPT promoting gambling. \label{fig:cpt_example_3}]
      {
        \includegraphics[width=.45\columnwidth]{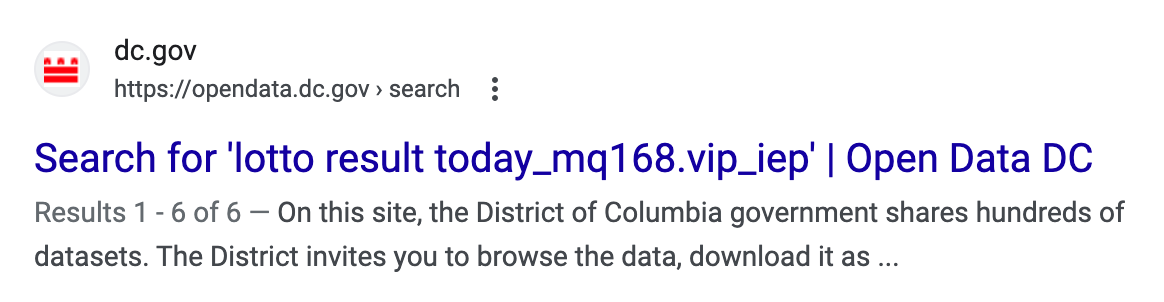}
    }
    \hfill
    \subfigure[An IPT promoting sex services in Korean. \label{fig:cpt_example_4}]{
        \includegraphics[width=.45\columnwidth]{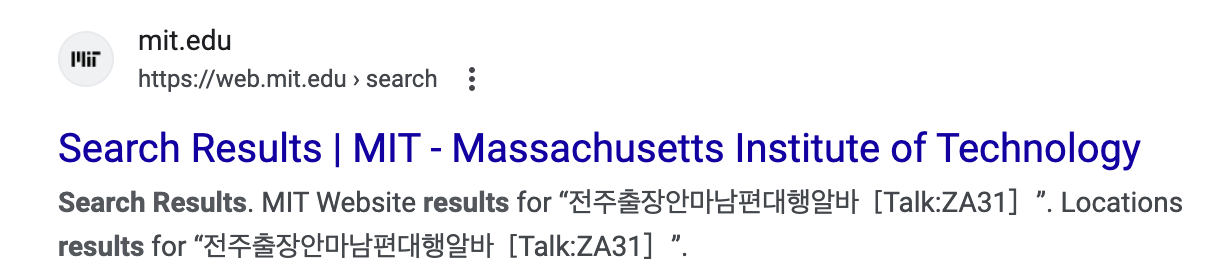}
    }

    \caption{Examples of IPTs as indexed by Google Search via reflected search poisoning.}
    \label{fig:cpt_examples}
\end{figure}

Given this well-crafted groundtruth, we have explored two text classification options. One is the combination of manually designed features and classic classification algorithms (e.g., Decision Tree and Random Forest), which features good explainability and high inference efficiency. The other is the fine tuning of a pre-trained large language model (LLM), which can avoid the burden of manual feature engineering and is known to have a higher data efficiency and a better generalizability.

When manually crafting features for the first classification option, 
we observed that IPTs tend to be much more lengthy than text samples from legitimate URL reflection cases. This is because a typical IPT contains three components: one or more popular search keywords, description of the services/goods under promotion, and contact information. Therefore, the number of characters turns out to be a good feature to distinguish IPTs from non-IPTs. Besides, benign URL reflection cases tend to contain only text characters while most IPTs were found to embed various non-text characters or pictorial symbols in order to highlight either the products under promotion or the contact information. Concrete examples include emojis (Fig.~\ref{fig:cpt_example_1}) and various brackets (Fig.~\ref{fig:cpt_example_4}). Therefore, extra features are designed to profile the existence and the count of such kinds of non-text symbols. Furthermore, most IPTs embed contacts for victims to further reach the underlying operators. And such contacts are either websites or accounts of various instant messaging platforms, e.g., Telegram, WeChat, QQ, WhatsApp, etc. On the contrary, a non-IPT URL reflection case seldom contains such a contact. Thus, the existence of one or more such contacts constitutes another subset of features, e.g., the number of URLs, and the number of IM marks (e.g., \textit{tg} for Telegram). And all four examples in Fig.~\ref{fig:cpt_examples} embed one or more such kinds of contacts. In total, we have designed seven groups of features to distinguish IPTs from benign URL reflection cases, and the full feature list can be found in Appendix~\ref{appendix:IPT_features}.

During training and evaluation, four different classic algorithms have been explored for the first classification option: Decision Tree, Random Forest, AdaBoost, and SVM. For the second classification option, the multilingual BERT~\footnote{https://huggingface.co/bert-base-multilingual-cased}\cite{DBLP:journals/corr/abs-1810-04805}, a well-adopted transformer-based encoder model,  is chosen as the pre-trained model for fine tuning.  Also, following the common practice, 80\% samples in the groundtruth are randomly sampled out for training, while the 20\% are held-out for testing. Detailed hyper parameters specific to the classification model can be found in Appendix~\ref{appendix:hyperparams_binary_ipt}. 

\begin{table}
    \centering
    \footnotesize
    \caption{The performance of different binary IPT classification models.}
    \label{tab:ipt_binary_performance}
    \begin{threeparttable}
        \begin{tabular}{ccccr}
            \toprule
            Model & Precision & Recall & F1 Score & Inference Speed~\tnote{1} \\
            \midrule
            BERT & 98.86\% & 98.63\% & 98.75\% & 18/267~\tnote{2}\\
            Random Forest & 95.34\% & 97.95\% & 96.63\% & 11667\\
            Decision Tree &  95.29\% & 96.81\% & 96.05\% & 13831\\
            AdaBoost & 94.41\% & 96.13\% & 95.26\% & 11510\\
            SVM & 94.39\% & 95.90\% & 95.14\% & 13457\\
            \bottomrule
        \end{tabular}
        \begin{tablenotes}
            \item [1] The unit is samples per second. 
            \item[2] The BERT model has been tested on both CPU and GPU, and the respective speed is 18/267 respectively.
        \end{tablenotes}
    \end{threeparttable}
\end{table}

Table~\ref{tab:ipt_binary_performance} lists a direct comparison among IPT classification models in terms of performance and inference efficiency. We can see that the fine-tuned BERT model has achieved the highest performance in terms of both recall and precision. On the other hand, among classic classification models using manually crafted features, Random Forest achieves the best performance with a recall of 97.95\% and a precision of 95.34\%. When it comes to inference efficiency, all these models are evaluated on the same physical server that is equipped with an Intel(R) Xeon(R) Silver 4214R CPU, 128GB memory, and two NVIDIA GeForce RTX 3090 GPUs. We can see that the fine-tuned BERT model, while excelling in performance, is subject to a much lower inference throughput at 18/237 samples per second on the CPU and GPU respectively, making it less suitable for applications requiring high-speed inference. We also evaluated the data efficiency of these classification options, i.e., the impact of the training data size on the performance of the binary IPT classification model. Fig.~\ref{subfig:data_efficiency_binary_ipt_precision} presents the precision performance of models trained upon different fractions of the whole training dataset, while Fig.~\ref{subfig:data_efficiency_binary_ipt_recall} presents the recall performance. As we can see that the fine-tuned BERT model consistently achieves the highest precision score across all data fractions, demonstrating its robustness to variations in the training dataset size. 

\begin{figure}
    \centering
    \subfigure[The performance in precision.]{
        \label{subfig:data_efficiency_binary_ipt_precision}
        \includegraphics[width=.46\columnwidth]{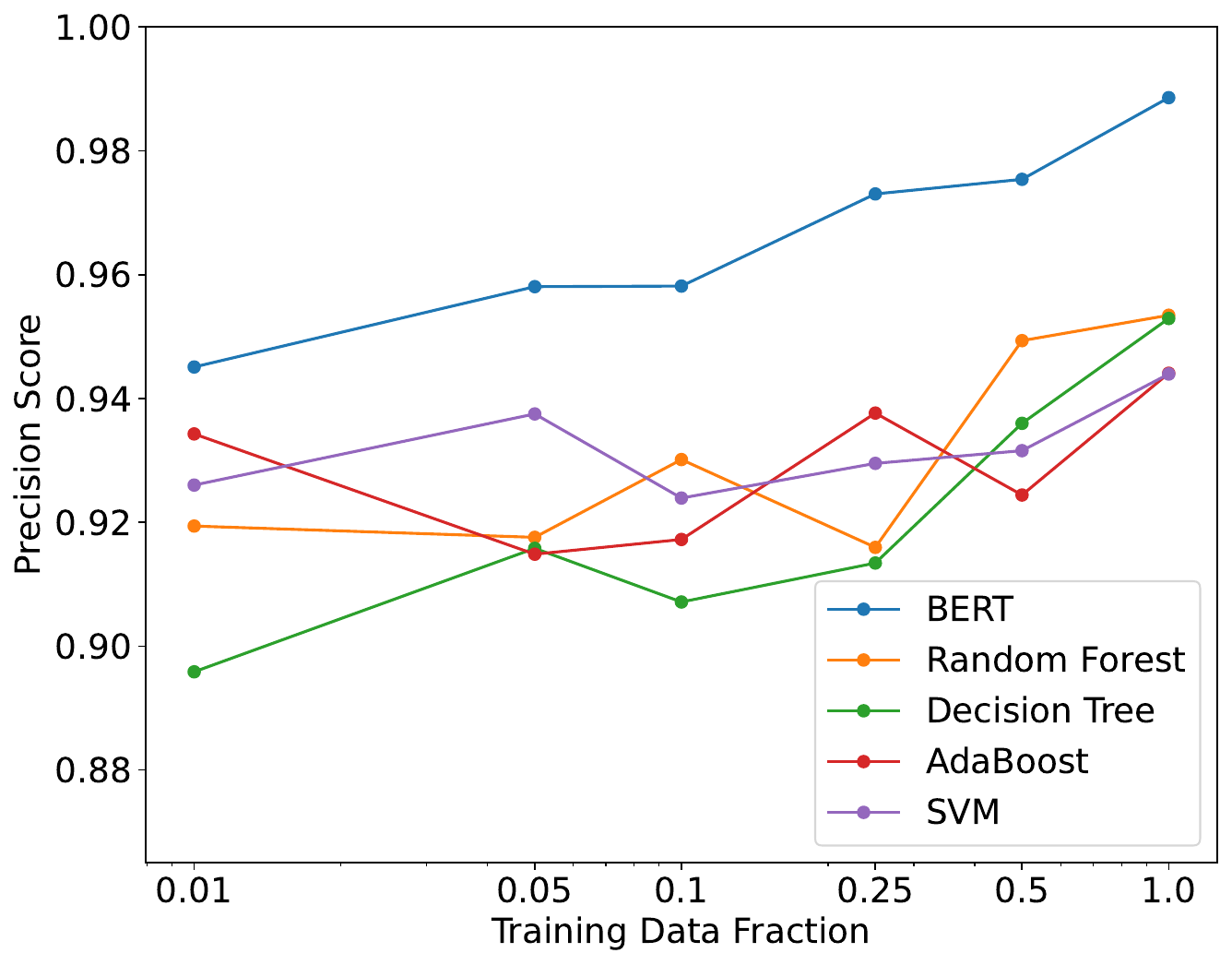}
    }
     \subfigure[The performance in recall.]{
        \label{subfig:data_efficiency_binary_ipt_recall}
        \includegraphics[width=.46\columnwidth]{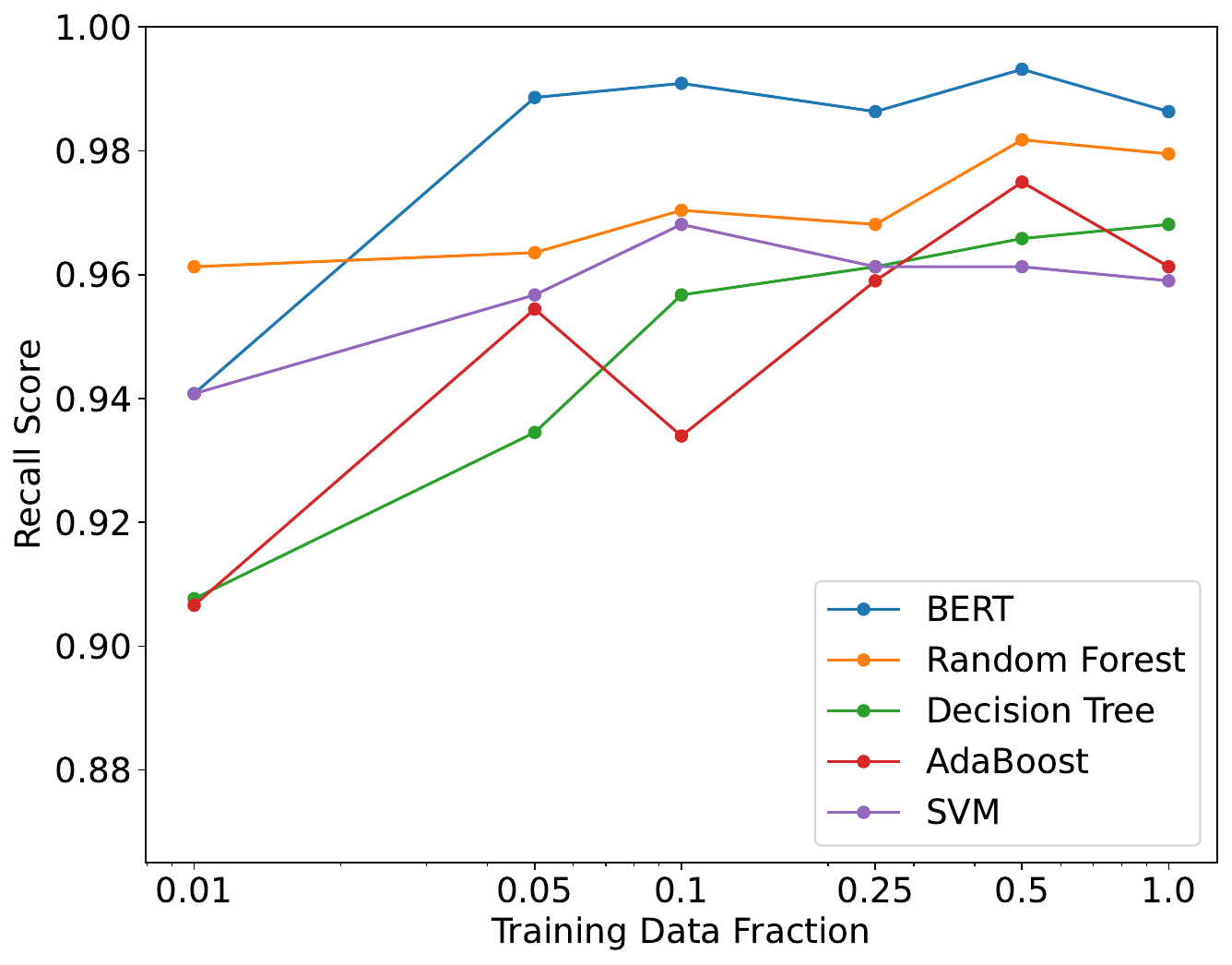}
    }
    \caption{The impact of training data size on the performance of binary IPT classifiers. }
    \label{fig:enter-label}
\end{figure}

Since this binary IPT classifier should be able to predict over 10M search result entries on a weekly basis, the Random Forest model is chosen as the default binary IPT classifier, as it achieves the best trade-off between prediction performance and inference efficiency, as revealed above. 



\subject{IPT keyword extractor}. 
Given newly captured IPTs, search keywords will be extracted to boost the next round of IPT hunting. As learned by case studies, the contact segment of an IPT turns out to be good search keywords in terms of guiding the search engines and discovering new RSPs/IPTs. 
As shown in Fig.~\ref{fig:cpt_examples}, the contact segment is usually separated from other IPT segments with non-text Unicode characters, e.g., brackets and pictorial symbols.
Therefore, to extract the contact segment out of an IPT, a regular expression is first applied to extract URLs, if any. Then, the left IPT text is further split into segments using a well-crafted list of separators including brackets (e.g., \{\}, [], 【】), pictorial symbols, and various punctuation marks. 
Next, a machine learning classifier is applied to decide whether a given IPT segment is a contact segment or not, and we name this classifier as the contact segment classifier. The resulting URLs (mostly domain names) and contact segments will be considered as IPT keywords.
 
 To build up this contact segment classifier,  8 features has been designed. Particularly, one feature is designed to capture the number of marks of instant messaging (IM) platforms since most contact segments contain one or more IM accounts. One more feature is the number of punctuation symbols, as such symbols are frequently used to separate an IM mark and the respective IM account identifier, e.g., the colon in "telegram:@ts775". The full list of features can be found in Appendix~\ref{appendix:contact_features}. Given the features designed, a ground truth dataset of 1,012 contact segments and 3,170 non-contact segments was collected to train and evaluate classification models. As a result, a random forest classifier has achieved the best performance of 90.51\% recall, and 91.05\% precision, which is satisfactory since this contact segment classifier is designed to identify good searching keywords and a false classification impacts only our searching efficiency.



\subject{Deployment}. 
When deploying our IPT hunter, a single round starts from querying search engines, and ends when new IPTs have been discovered along with new IPT keywords and new URL reflection schemes extracted. In the search step of each round, our crawlers strictly follow the rate limit policies of search engines and will stop when an HTTP response error indicating the rate limit is reached, and only restart when the rate limit is cleared up.
Therefore, to speed up the discovery process, we subscribed to multiple web proxy services and relayed our search engine queries through their globally distributed proxy servers. 
Regarding which search engines to query, two general and popular search engines were first selected, namely Google Search and Microsoft Bing. Also, considering that many Chinese IPTs were observed during our preliminary study, two more Chinese search engines, namely Baidu Search and Sogou Search, were selected.

Then, to jump-start the crawler, we first manually crafted a seeding set,  
\arevised{
which involves an iterative process of searching Google with existing IPT keywords, manually labeling searching results and extending the IPT keyword set. 
}
\arevised{
This began with a Google query \textit{site:https://ctan.org/search}, which is a poisoned in-site search of an English-language repository containing TEX-related materials for download. We then parsed the search results to extract relevant keywords. These keywords were used to perform additional Google searches, expanding the pool of potential URL reflection schemes, and these schemes were utilized to search for more keywords in turn. This iterative process was repeated over four cycles. We manually classified the acquired results to ensure relevance and reliability. Ultimately, we successfully obtained a seeding set comprising 153,385 IPT keywords and 15,113 URL reflection schemes, covering various categories and languages.} Leveraging this seeding set, we then applied this hunter to all four search engines and completed four rounds between Nov. 8, 2022 and Dec. 5, 2022. Then, considering the high expense of web proxies, we suspended the IPT hunter for all search engines except Bing, for which our IPT hunter continued to run until Apr. 1, 2023 thanks to its lax rate limit. 
    We name these rounds as the first-phase IPT hunting. Then, to profile the longitudinal evolution of IPTs, we conducted a second-phase hunting which involves seven rounds during the time period between Nov. 6, 2023 and Dec. 27, 2023.
In total, we have discovered 13,295,628 distinct RSP incidents, 11,957,205 distinct IPTs and 180,757 URL reflection schemes, for which more details will be presented in \S\ref{sec:promotions}.

Besides, the deployment has well demonstrated the snowballing effectiveness of our IPT hunter. For instance, the URS \textit{https://www.ze-dge.net/find/$KW$} is designed to search available wallpapers matching the given keyword $kw$. Given this URL reflection scheme,  our hunter has identified 6,810 IPTs that promote illicit services/goods of 13 different categories. On the other hand, searching with the IPT keyword \textit{cheapfifa23coins.com} allowed us to identify 1,784 IPTs that had abused 1,582 different URSes. 

\subsection{The IPT Analyzer}
\label{subsec:method_cpt_analyzer}
Given the large volume of IPTs captured, two additional tools have been developed to automatically analyze IPTs and gain a deep understanding of RSP-based illicit promotion. One is a multi-label classifier which groups an IPT into one or more services/goods categories, e.g., counterfeit goods, and fake certificate services. Furthermore, a contact extractor is designed to extract from a given IPT,  contacts of various types including websites, instant messaging accounts, and telephone numbers.

\subject{The multi-label IPT classifier}. 
 This classifier is designed to profile what categories of services/goods an IPT is intended to promote. It takes an IPT as the input and classifies it into one or more categories. By manually labelling randomly sampled IPTs,  14 categories have been defined for different services/goods, and the full list can be found in Table~\ref{tab:cybercrime_categories}.
 Regarding the naming of these IPT categories, we adopt some terms proposed in previous works~\cite{yang2017learn, yuan2018reading,180611,lin_fake_2011,8121871,west_intelligent_2016}, such as Gambling, Data Theft and Fake Certificate. Then, the legitimacy of some services/goods categories can vary significantly from one jurisdiction to another. For instance, many IPTs promote surrogacy services, sex services, or drug sales in Chinese, all of which however are illegal in mainland China.  When naming such categories, we take an objective standard and try to avoid implying the legitimacy in the names. 
 Also, we consider multilingual texts as the input since IPTs are observed to be multilingual. 

\begin{table}\scriptsize
    \caption{The list of categories of goods and services.}
    \label{tab:cybercrime_categories}
    \centering
    \setlength{\tabcolsep}{7mm}{
    \begin{threeparttable}
        \begin{tabular}{cc}
            \toprule
            Category & Category\\
            \midrule
            Black Hat SEO \& Advertisement & Hacking Service\\
            Counterfeit Goods &  Drug Sales\\
            Data Theft & Sex Service\\
            Fake Account & Surrogacy Service\\
            Fake Certificate &  Weapon Sales\\
            Financial Fraud & Money Laundering\\
            Gambling & Others\tnote{1}\\
            
            \bottomrule
        \end{tabular}
        \begin{tablenotes}
            \footnotesize
            \item[1] Categories like ghost writing and detective employment are grouped as "Others" due to their small sample size.
        \end{tablenotes}
    \end{threeparttable}
    }
\end{table}

To build up this classifier,
a ground truth dataset of 7,246 IPT samples were labeled. Then, regarding the classification algorithm,  we chose to fine tune the aforementioned multilingual BERT model, due to the following considerations. First of all, the fine-tuned BERT has achieved the best performance along with good data efficiency in the binary IPT classification task. Also, different from the binary IPT classifier, this multi-label one is used for offline analysis for which the inference speed is not a bottleneck. Through training on 80\% of the ground truth dataset, the resulting model has achieved good performance when evaluated with the held-out 20\% ground truth dataset. Particularly, the micro precision of the resulting model is 94.03\%, while the micro recall is 93.53\% and the LRAP (label ranking average precision) score is 0.9593. The per-class performance metrics can be found in Appendix~\ref{appendix:performance_cybercrime_classifier}.




\subject{The IPT contact extractor}. Taking an IPT as the input, our contact extractor is designed to extract all the embedded contact entities, which is achieved through three major steps.
The first step is a preprocessing step, which is intended to remove various noise added by miscreants likely in an attempt to impede automatic extraction, such as the usage of homoglyphs and splitting a domain name with human-readable characters. 
Then, the second step aims to decide the type of embedded contact. Then, depending on the contact type, either a named entity recognizer (NER) or a rule-based extractor will be applied to identifying the respective contact entity. 

\subsubject{The pre-processing step}. As revealed by our case studies, the operators of IPTs adopt various evasion tactics when embedding contacts. Although the resulting contacts are still human-readable, it is challenging to get them automatically extracted. Typical evasion tactics include the replacement of alphanumeric characters with respective homoglyphs and replacing the dot separator in a domain name with a Chinese period symbol. Therefore, in this pre-processing step, such kinds of evasion noise will first be removed.

\subsubject{The contact type classifier}. 
This classifier takes an IPT as the input and decides which type of contact it contains. Currently, this classifier supports the classification of 5 contact types that are most frequently embedded in IPTs. These contact types include telephone numbers, Telegram accounts, WeChat accounts, QQ accounts, and websites. And an IPT will be classified as "others" if it doesn't contain any contacts that belong to those 5 types.
To build up this classifier, the aforementioned transformer-based multilingual language model is reused and further tuned on a ground truth dataset of 5,482 labeled IPTs, and the resulting model has achieved a micro precision of 94.91\% and a micro recall of 94.91\%.

\begin{table}
    \centering
    \footnotesize
        \caption{The performance of classification models for contact extraction.}
    \label{tab:contact_performance}
    \begin{tabular}{crrr}
        \toprule
        Model Name & Precision & Recall & F1-Score \\
        \midrule
        The contact type classifier & 94.91\% & 94.91\% & 94.91\%  \\
        The WeChat NER model & 96.22\% & 98.71\% & 97.45\%\\
        The Telegram NER model & 98.15\% & 100.00\% &99.07\%\\
     \bottomrule
    \end{tabular}
\end{table}

\subsubject{Contact entity extractors}.
Given the contact type decided for each IPT, the next step is to extract the respective contact entity. For contact types of websites, telephone numbers, and QQ accounts, simple rule-based extraction works well thanks to their strict format, e.g., both telephone numbers and QQ accounts are numeric strings. For the extraction of Telegram and WeChat accounts, we consider them as tasks of named entity recognition (NER). As transformer-based language models~\cite{yan2019tener, li2020flat} have also achieved state-of-the-art performance in many NER tasks. We utilized the aforementioned multilingual BERT model to build the two separate NER models for recognizing WeChat accounts and Telegram accounts respectively.
Both NER models have achieved good performance, as detailed in Table~\ref{tab:contact_performance}.

As a result, we have identified 48,114 distinct contact entities in total, which consist of 16,335 websites, 5,890 Telegram accounts, 23,632 WeChat accounts, 1,552 QQ accounts, and 705 telephone numbers.







\subsection{The IPT Infiltrator}
\label{subsec:method_cpt_infiltrator}

The contact entities identified above serve as the next hops for victims (potential customers) to interact with operators of the promoted goods or services. Infiltrating these contacts can help gain a better understanding for the underlying illicit campaigns. Therefore, an IPT infiltrator is designed and implemented to fulfill this goal. 

\subject{The dynamic website crawler.} 
For website contacts, we periodically visit each of them at a weekly pace through instrumenting a headless browser. We capture the final landing webpage as a screenshot and save all the network traffic of both HTTP requests and HTTP responses. Also, if a landing page is found to be promoting a mobile app, it will trigger a further inspection during which we manually download the respective mobile app for further threat analysis. 
Specifically, this dynamic website crawler is built upon the Playwright framework~\footnote{https://playwright.dev} which offers APIs in many programming languages and enables end-to-end automatic testing of websites.

\subject{The Telegram account infiltrator.} Besides, leveraging publicly available Telegram APIs~\footnote{https://core.telegram.org/}, the profile of each Telegram account will be weekly extracted. Then, if an account turns out to be a Telegram channel, our infiltrator will subscribe to the channel and retrieve all its historical messages since 2022 and weekly obtain its new messages. Similarly, when an account is found to be a Telegram group, our infiltrator joins the group, weekly retrieves the latest group members and messages, and downloads all the historical messages since 2022. Furthermore, among messages collected from either Telegram channels or groups, many of them are used to promote other Telegram channels or groups. For such kinds of messages, our infiltrator further snapshots the involved Telegram groups or channels by following a similar infiltration strategy.

Different from Telegram, neither WeChat nor QQ has any public APIs to facilitate programmable access to account profiles or group messages. Instead, what we can do is to manually search each account and view its public profile. Therefore, we conducted manual inspection on a  randomly sampled set of 250 WeChat accounts and 200 QQ accounts in an attempt to learn their activeness, locate illicit promotion indicators if any, and understand the tactics underpinning their escape from being blocked from the respective platform. The detailed results of our infiltration will be presented in \S\ref{sec:infra}.


\subsection{Ethical Considerations}
\label{subsec:ethical}
When designing and implementing the methodology, we have taken necessary measures to minimize any potential ethical issues. When collecting data from search engines and instant messaging platforms, we have respected their corresponding user policies, and carefully limited the crawling rate to avoid any non-negligible burden to the respective platform. Also, all the collected raw datasets are securely stored on our research servers and the data access is only granted to our researchers. One thing to note, some websites of IPTs were later found to contain child pornography content, for which, we have completely removed the media files as retrieved from these websites.

\section{RSP-based Illicit Promotion}
\label{sec:promotions}

In this section, we present a comprehensive measurement for illicit promotion texts (IPTs) that are distributed through reflected search poisoning (RSP), i.e., RSP-based IPTs.  As a result, a deep understanding is gained, for the first time, regarding  what services/goods have been illicitly promoted through RSPs (\S\ref{subsec:landscape}), how IPTs evolve across time,  what benign websites have been abused (\S\ref{subsec:abused_websites}), as well as the extent to which benign search users can be exposed to IPTs. (\S\ref{subsec:target_of_CPT}).  


\subsection{Illicit Promotion Texts}
\label{subsec:landscape}

\begin{table}
  \centering
  \footnotesize
  \caption{The scale statistics of RSP-based IPTs.}
  \label{tab:scale_cpts}
  \resizebox{\linewidth}{!}{
  \begin{threeparttable}
    \begin{tabular}{crrrr}
      \toprule
      Group & IPTs & RSPs & URL Schemes & Abused FQDNs\\
      \midrule
      Google &  11,768,050 & 13,060,671 & 170,828 & 74,985\\
      Baidu & 68,883 & 90,220 & 4,988 & 2,706\\
      Sogou & 6,802 & 7,658 & 2,594 & 1,417\\
      Bing & 459,333 & 490,285 & 34,701 & 14,369\\
      Overall & 11,957,205 & 13,295,628 & 180,757 & 79,317\\
      \bottomrule 
    \end{tabular}
  \end{threeparttable}
  }
\end{table}

\subject{Scale}. 
In total, we have captured 11,957,205 distinct IPTs that have been distributed through 13,295,628 RSPs. These RSPs have abused 180,757 unique URL reflection schemes, which belong to 79,317 fully qualified domain names (FQDNs) and 60,638 apex domains.
Table~\ref{tab:scale_cpts} presents more detailed scale statistics for IPTs. And we can see IPTs via RSPs have successfully poisoned all four search engines under our study at a large scale.
Among the four search engines, the two Chinese search engines, namely Baidu and Sogou, have much fewer IPTs than the other two global counterparts, which is likely because these two Chinese search engines have already enforced effective measures to filter out IPTs and RSPs.

 However, the volume of IPTs is not equivalent to that of the underlying promotion campaigns, since the same campaign can promote its services or goods through many different IPTs. Instead, we observe that IPTs belonging to the same campaign tend to share a common set of very few contacts. Leveraging the IPT contact extractor (\S\ref{subsec:method_cpt_analyzer}), we have observed 48,114 distinct contacts, which suggests that the scale of the underlying promotion campaigns is also very large.  

\begin{table}
  \centering
  \footnotesize
  \caption{The distribution of IPTs over categories of goods and services.}
  \label{tab:cpt_categories}
    \resizebox{\linewidth}{!}{
    \begin{threeparttable}
    \begin{tabular}{cr|cr}
    \toprule
    Category & \% IPTs & Category & \% IPTs\\
    \midrule
    Sex Service & 25.39\% & Drug Sales & 2.29\%\\
    Gambling & 23.72\% & Surrogacy Service & 1.93\%\\
    Fake Certificate & 22.65\% & Others & 1.32\%\\
    Black Hat SEO \& Advertisement & 9.16\% & Counterfeit Goods & 1.26\%\\
    Fake Account & 4.79\% & Financial Fraud & 1.09\%\\
    Hacking Service & 2.75\% & Money Laundering & 0.77\%\\
    Data Theft & 2.72\% & Weapon Sales & 0.16\%\\
    \bottomrule
    \end{tabular}
    \end{threeparttable}
  }
\end{table}

\subject{Categories.} 
Using the multi-label text classifier ( \S\ref{subsec:method_cpt_analyzer}), we categorize captured IPTs into 14 categories of services and goods. As shown in Table~\ref{tab:cpt_categories}, the top categories with most IPTs include Sex Service (25.39\%), Gambling (23.72\%), and Fake Certificate (22.65\%). Notably, black hat SEO operators and other illicit advertisement services have also used RSP  to promote their services, accounting for 9.16\% of IPTs. 
Also, regarding the legitimacy of the promoted services and goods,  over 45\% IPTs are used to promote categories of service and goods that are commonly considered as illegal across different jurisdictions, e.g., Fake Certificate, Hacking Service, Financial Fraud, and Counterfeit Goods. 
On the other hand, other categories vary significantly 
 in their legitimacy across different jurisdictions. For instance,  many states in the United States consider gambling of some types as legal, while gambling in mainland China is forbidden. Besides, sex services (prostitution) are legal in Netherlands, while it is illegal in the United States except for Nevada. Therefore, IPTs under these categories are inherently controversial, which partly explains why the underlying operators of such services/goods choose RSP-based illicit promotion rather than legitimate promotion channels (e.g., Google Ads).


To understand which specific products are being promoted for each category, we examined 500 IPTs in each category to identify specific products. Particularly, in the category of Fake Account, fraudulent accounts of 20 different online platforms were for sale, e.g., Amazon, Twitter, Alipay, PayPal, Uber, and WeChat.  Besides, 
In the category of Fake Certificate, we identified counterfeit services for 19 types of documents, e.g., identification cards and passports, diploma and transcripts of top universities, medical records, etc. Then, the category of Hacking Services encompassed not only services for DDOS attacks and domain hijacking, but also services of developing fraud platforms and gambling websites. 
Regarding the category of Drug Sales, 13 different illegal drugs were promoted,  e.g., methamphetamine, sleeping pills and psychedelics. Furthermore, for the category of Money Laundering, our investigation revealed that criminals were trying to collect large numbers of bank cards and hire regular users to assist in money laundering. 
Please refer to Appendix~\ref{appendix:IPT_products} for typical products of all the categories.

\subject{Natural languages}.
We further profiled the IPTs with regards to their natural languages, wherein the natural language of each IPT is identified through a language identification tool named \textit{langid}~\footnote{https://github.com/saffsd/langid.py}. In total, we have observed 97 natural languages, with the top five being Chinese (88.08\%), Korean (4.86\%), English (1.66\%), Japanese (1.48\%), and Vietnamese (0.95\%).
As we can see,  this language distribution of IPTs is contrasted with that of the whole Internet~\cite{internet_language} wherein the English accounts for almost half of all the web content. We then looked into the intermediate results of our IPT hunter and have thus confirmed the fidelity of this language distribution. 

And we believe several factors can jointly contribute to this skewed language distribution. One is that the reflected search poisoning may been utilized more heavily by illicit promotion campaigns targeting regions of East Asia and Southeast Asia, which can lead to more IPTs composed in natural languages of these regions. The other explanation is that search engines especially Google and Bing vary a lot in their capabilities to filter out IPTs of different languages, and IPTs in CJK languages have a higher probability to bypass the filtering and get indexed by these search engines, which leads to the survivorship bias.
We also compared IPTs of different natural languages in terms of their category distribution, for which more details can be found in Appendix~\ref{appendix:category_distribution}.





\subject{Availability across search engines}.
When it comes to the distribution of IPTs across search engines, an interesting observation is that IPTs observed on one search engine can be mostly exclusive to ones captured from another search engine, likely due to different indexing and filtering strategies of different search engines. For instance, the IPT overlap rate of Bing, Baidu, and Sogou across Google is only 13.54\%, 17.65\%, and 3.33\% (Bing, Baidu, Sogou as the denominator), respectively. Further details can be found in Appendix~\ref{appendix:cross_SE_IPT_distribution}.

\begin{table}
    \centering
    \footnotesize
    \caption{The temporal evolution of IPTs with regards to scale.}
    \label{tab:evolution_ipt_scale}
    \begin{tabular}{crrrr}
        \toprule
        \multirow{2}{*}{Group} &  \multicolumn{2}{c}{Nov, 2022} & \multicolumn{2}{c}{Nov, 2023}\\
        \cmidrule(r){2-3}\cmidrule(r){4-5}
         & IPTs & Contacts & IPTs & Contacts\\
         \midrule
         Google & 6,917,708 & 20,224 & 5,112,992 & 31,981\\
         Baidu & 66,595 & 2,024 & 25,150 & 1,521\\
         Sogou & 6,968 & 316 & 2,376 & 158\\
         Bing & 458,710 & 5,050 & 672 & 132\\
         Overall & 7,085,944 & 22,905 & 5,137,488 & 32,596\\
         \bottomrule
    \end{tabular}
\end{table}

\subject{The longitudinal evolution.} 
To profile the temporal evolution of IPTs, we deployed the IPT hunter in November 2022 and then in November 2023, which gives two separate IPT datasets spanning a time gap of around 12 months. Comparing the two separate IPT datasets reveals a set of interesting observations. First of all, by November 2023, RSP-based illicit promotion is still comparable to that of November 2022, with regards to scale and diversity. 
However, we also observe that 94.82\% of the IPTs identified in 2023 were newly captured, not observed during the first-phase in November 2022. Similarly, 77.34\% of IPT contacts were new, highlighting the quick evolution and  adaptability of illicit promotion. This dynamic turnover also suggests an ongoing cat-and-mouse game between regulatory efforts and operators of illicit promotion.

As shown in Table~\ref{tab:evolution_ipt_scale}, all the search engines except for Bing still suffer from a similar extent of IPT poisoning. However, for Bing, the number of IPTs has dropped from over 458K in November 2022 to almost zero in November, 2023. One thing to note, we conducted responsible disclosure to Bing in October, 2023 and Bing responded that they are working to address this issue. We thus believe the vanishment of IPTs is likely due to the mitigation measures enforced by Bing since our disclosure. 

When it comes to the evolution of IPT categories, some categories have seen obvious increase in the ratio of IPTs that each of them accounts for. For instance, the ratio of IPTs belonging to Drug Sales has increased from 1.21\% in 2022 to 3.95\% in 2023 while the category of Data Theft has seen an increase from 1.60\% to 4.47\%. On the contrary, the category of Gambling has seen a drop in the IPT share from 26.52\% to 19.79\%. This is likely because the FIFA World Cup was held in 2022 while there is no comparable sports event in 2023, which leads to less illicit promotion of gambling services in 2023. 

\subsection{Websites Abused in Illicit Promotion}
\label{subsec:abused_websites}
We then profile the high-ranking websites that have been abused in RSPs for distributing IPTs.

\subject{Scale and distribution}. In a nutshell, a large volume of high-ranking websites has been abused in RSP attacks while each abused website can be free rode by IPTs belonging to different promotion campaigns. Specifically, 
180,757 distinct URL reflection schemes (URSes) have been abused in one or more RSPs, which belong to 79,317 fully qualified domain names (FQDNs) and 60,638 apex domains. Examples of mostly abused URSes and apex domains are listed in Appendix~\ref{appendix:top_abused_website}. Also, a long-tailed distribution of IPTs over URSes is observed, as the top 5\% URSes have accounted for 87.16\% RSPs and 87.94\% IPTs while they are 95.52\% RSPs and 95.91\% IPTs for the top 10\% URSes. 

Furthermore, one domain name may have up to hundreds of URSes abused in reflected search poisoning.  For example, \textit{baidu.com}, an apex domain owned by the same company operating Baidu Search, has 270 URL reflection schemes being abused by miscreants, e.g., \textit{http://zhidao.baidu.com/q?XX} (among the top 20 mostly abused URL schemes), and \textit{https://ai.baidu.com/for-um/es/search}. The famous video website YouTube also have 29 URSes abused, e.g., the in-site search page (\textit{https://music.youtube.com/search}) and the hashtag page (\textit{https://youtube.com/hashtag}). 

\begin{table}
    \caption{The abuse of most popular websites in RSP attacks.}
    \label{tbl:tranco}
    \centering
    \footnotesize
    \begin{tabular}{lrrr}
    \toprule
            Group &  Abused Domains &  \% IPTs &  \% RSPs\\
    \midrule
        Top 100 &                       46 &          3.70\% &       3.48\% \\
       Top 1K &                      364 &         9.69\% &       9.28\% \\
      Top 10K &                     2,113 &         21.59\% &       20.53\% \\
     Top 100K &                     8,006 &         42.96\% &       40.54\% \\
    Top 1M &                    20,330 &        67.46\% &       63.53\% \\
    \bottomrule
    \end{tabular}
\end{table}


\subject{Reputation and popularity.} One more interesting question is regarding whether websites abused in RSPs indeed have a good reputation, so as to increase the likelihood of getting their webpages (including RSPs) indexed by search engines with high page ranks. To answer this question, we use the metric of website popularity as an approximation to website reputation. To profile the popularity of each abused website, we referred to Tranco~\cite{Tranco}, a top site ranking with 1 million websites listed. As a result, we observed that many abused websites are highly popular. Specifically, as listed in Table~\ref{tbl:tranco}, among the aforementioned 60,638 apex domains, 20,330 (33.53\%) show up in the top one million, 2,113 (3.48\%) in the top 10K, and 46 (0.08\%) even rank in the top 100.
Besides, these abused top websites contribute  most of the IPTs. Particularly, 67.46\% IPTs were observed from websites in the top one million. Also, 8,006 of the top 100K websites, while accounting for only 13.20\% of all abused websites, have contributed 42.96\% IPTs and respective 40.54\% RSPs.


\subject{The functions of URL reflection schemes}. 
What is also found is that these abused URL schemes are intended for many different but all legitimate usage scenarios. And 3 typical usage scenarios include onsite search, hashtag, and web dictionary or translation. Specifically, among the top 100 most abused URL reflection schemes, 80\% are intended for onsite search, 11\% are designed for hashtags, and 6\% are for web dictionary or translation. 
There are also some other types of URL schemes, for example, \textit{https://minecraft.fandom.com/zh/wiki/XX} is a wiki page for "XX". However, the website may still reflect the content of the URL even if "XX" is not a valid term in the wiki.

\begin{figure}
    \centering
    \subfigure[
        A Chinese keyword searching for a city name (Fengtai, Beijing).
        \label{fig:CPT_target_case_2}
    ]{
        \includegraphics[width=.45
        \columnwidth]{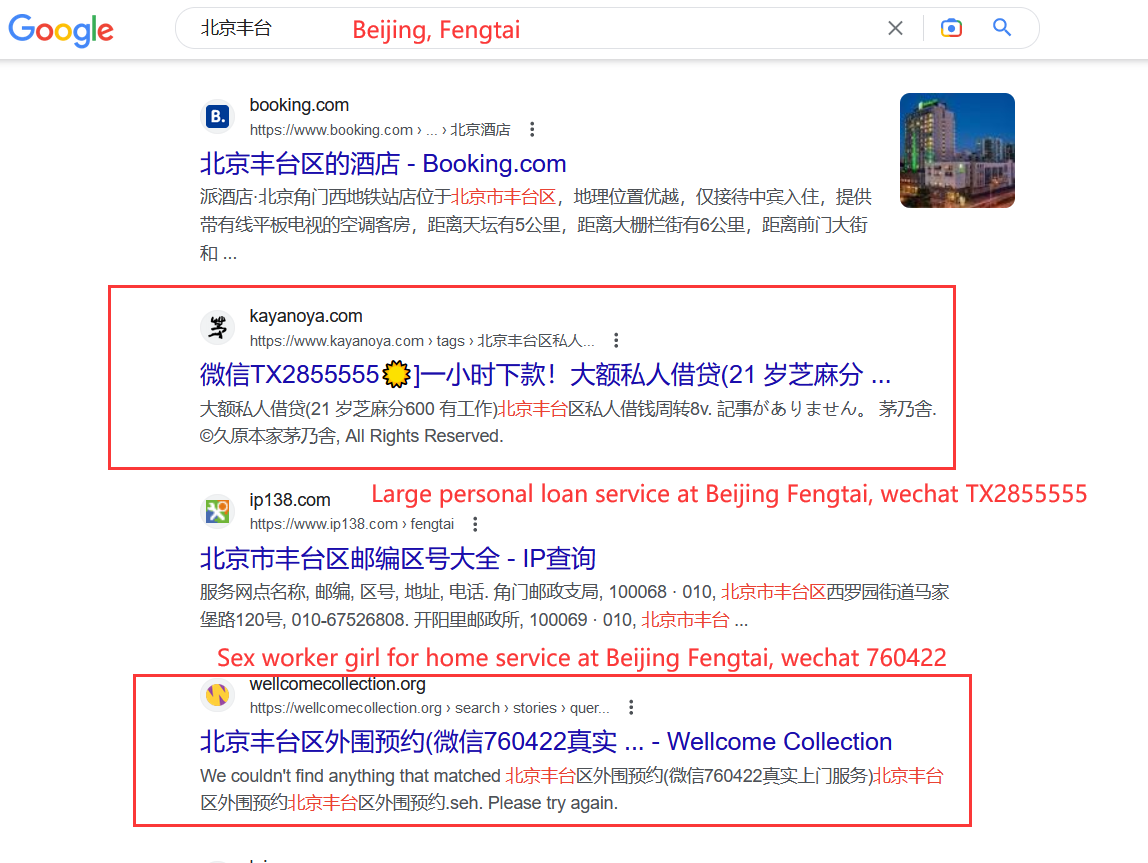}
    }
    \hfill
    \subfigure[
        A Chinese keyword searching for sex services. 
        \label{fig:CPT_target_case_1}
    ]{
        \includegraphics[width=.45
        \columnwidth]{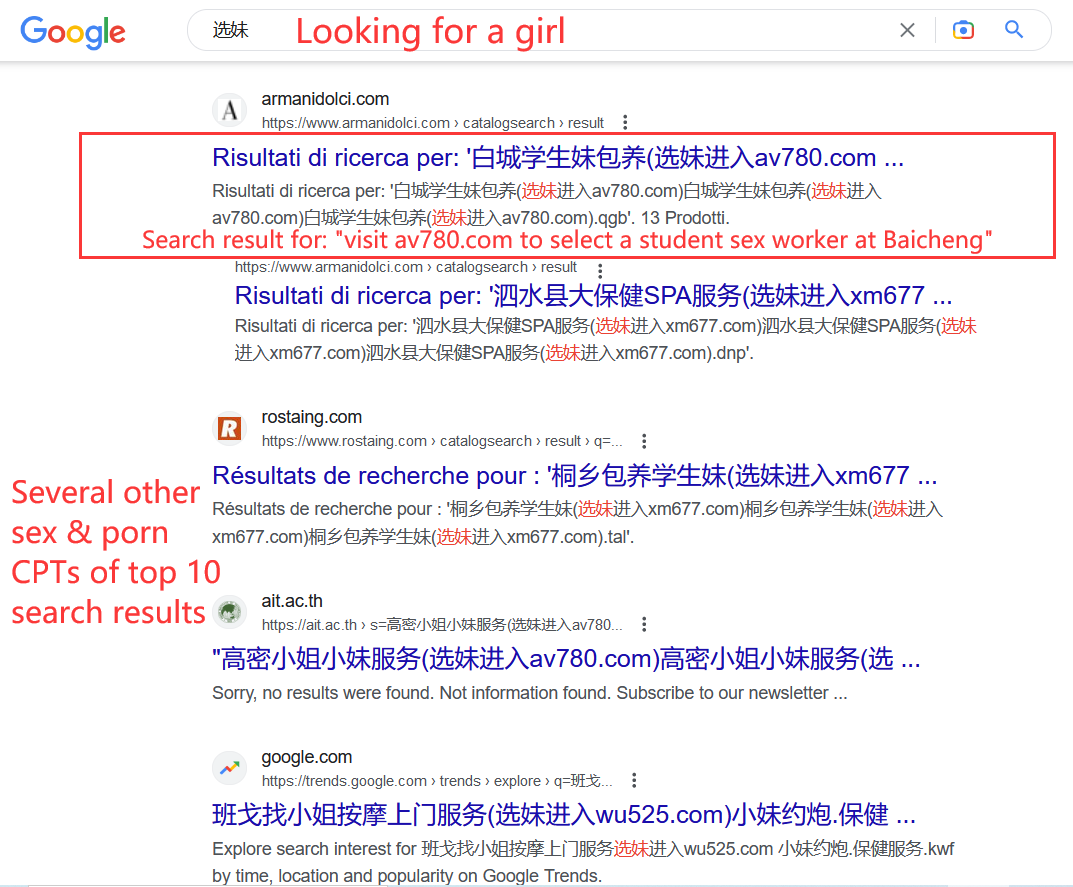}
    }

    \subfigure[
        A Chinese keyword searching the price of Dior women's athletic shoes. 
        \label{fig:CPT_target_case_3}
    ]{
        \includegraphics[width=.45
        \columnwidth]{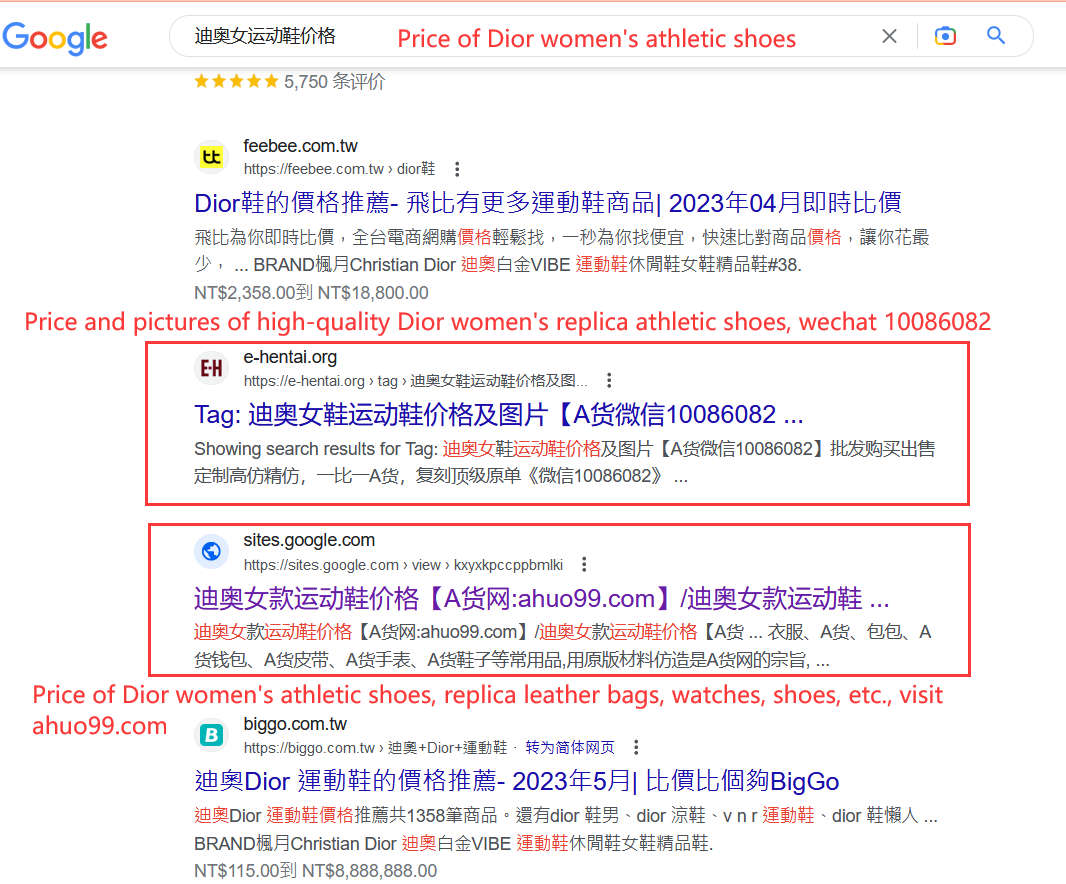}
    }
    \hfill
    \subfigure[
         A Chinese keyword searching the train from Taizhou to Huanggang.
        \label{fig:CPT_target_case_4}
    ]{
        \includegraphics[width=.45
        \columnwidth]{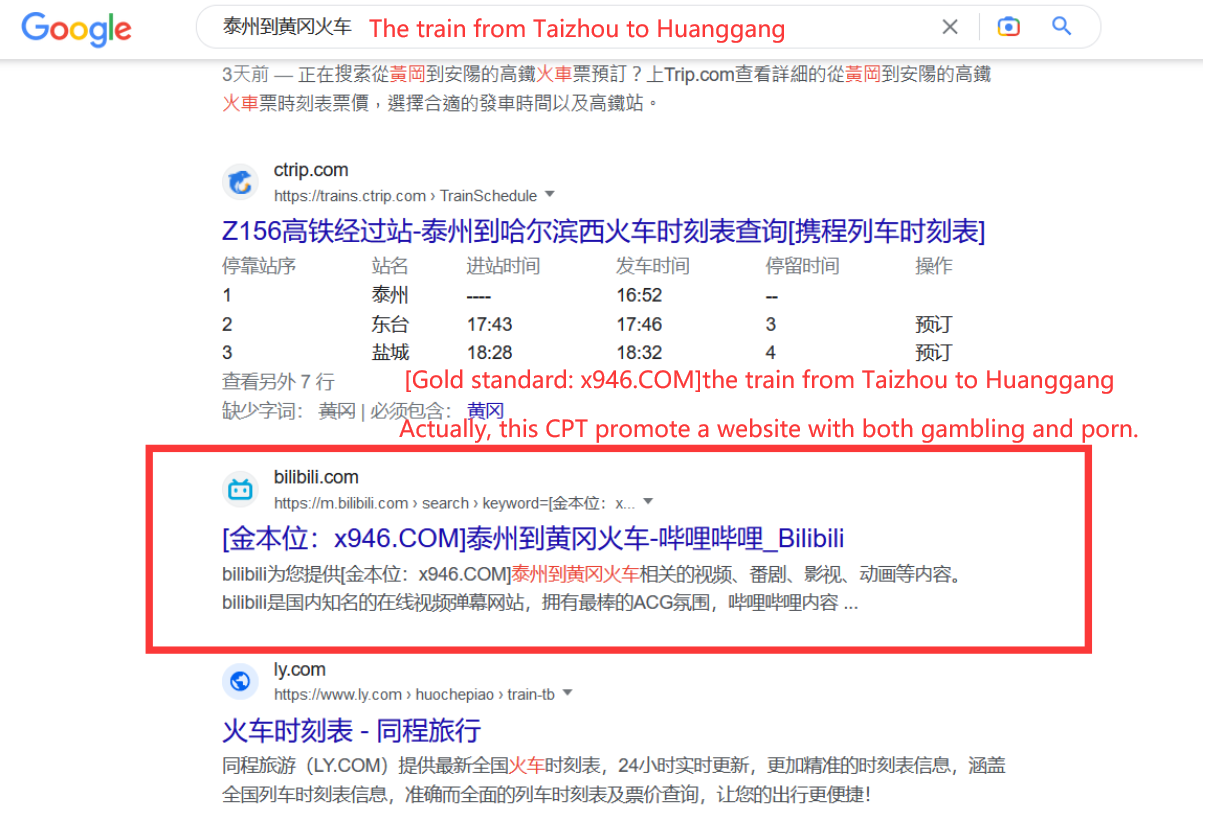}
    }
    \caption{The cases of search keywords of different categories that will likely expose search user to IPTs. }
    \label{fig:CPT_target}
\end{figure}

\subsection{The Exposure of Search Users to IPTs}
\label{subsec:target_of_CPT}
With search engines poisoned with so many IPTs, it is intuitive to ask: \textit{to what extent can search users be exposed to IPTs?} To answer this, we manually studied IPT cases and performed several search experiments. As a result, we have identified three groups of search keywords that are commonly used by regular search users and may expose search users to IPTs:  location names (e.g., city names), keywords of services and goods especially illicit ones, and benign long-tail keywords.

The first type is keywords denoting location names especially city names. When we search for some location names, especially the city name, search engines are likely to return IPTs embedding these location names. For example, when searching in Chinese for \textit{Fengtai, Beijing} on Google, the searching results contain several IPTs promoting either loan services or sex services, as shown in Fig.~\ref{fig:CPT_target_case_2}. 
To further investigate this phenomenon, we used 3,368 Chinese city names to query four search engines and found that Google and Bing have been heavily poisoned, whereas Baidu and Sogou were not, possibly due to their extra efforts to filter such poisoned results. As shown in Table~\ref{tbl:city_name}, when querying Google with these city names, 1,557 out of 3,368 queries (46\%) have been poisoned with one or more IPTs in the top 10 result entries, and it is 94\% for the top 50 result entries. 

\begin{table}
\caption{The availability of IPTs in top search result entries when querying Google and Bing with Chinese city names.}
\label{tbl:city_name}
\centering
\footnotesize
\begin{tabular}{@{\extracolsep{4pt}}crrrr}
\toprule
        \multirow{2}{*}{
        Top Entries} 
        &  \multicolumn{2}{c}{Google} & \multicolumn{2}{c}{Bing}\\
        \cmidrule(r){2-3}\cmidrule(r){4-5}
        & \% Poisoned & IPTs  & \% Poisoned & IPTs\\

\midrule
        10 & 46.23\% & 4.7K & 0.42\% & 35 \\
    20 & 86.85\% & 20.8K & 0.45\% & 42 \\
    50 & 94.24\% & 70.2K & 0.68\% & 64 \\
    100 & 95.61\% & 112.7K& 3.89\% & 311 \\
    All & 95.96\% & 124.1K & 5.73\% & 437 \\
\bottomrule
\end{tabular}
\end{table}

The second type encompasses keywords of services and goods especially illicit ones, which is straightforward and easy to understand. For instance, if searching "fifa 23 coins in norway" on Google, we could get several IPTs such as "Visit Twitter.com/dcdgame| No worries at all when you buy best way to get coins fifa 23 in Norway!" on the first page. More directly, as shown in Fig.~\ref{fig:CPT_target_case_1}, if we search in Chinese for \textit{Looking for a girl} on Google, the first page is filled with Chinese IPTs related to sex services.

The third type describes that if we search for some benign long-tail keywords, the search engine may return IPTs embedding those keywords. For example, as shown in Fig.~\ref{fig:CPT_target_case_3}, if we search in Chinese for the price of Dior female athletic shoes on Google, it returns several IPTs about counterfeit shoes. Obviously, such IPTs are embedded with benign long-tail keywords and are used to promote related illicit services/goods. In another interesting case as shown in Fig.~\ref{fig:CPT_target_case_4},  when searching in Chinese for the train from Taizhou to Huanggang, Google returns an IPT promoting a website of gambling and porn videos. However, the long-tail keyword and the theme of this IPT are completely unrelated. And many IPTs have been found to embed long-tail keywords unrelated to their themes, likely in an attempt to reach a much broader search audience.









\section{Next Hops of Illicit Promotion}
\label{sec:infra}

In this section, we report an in-depth analysis of contacts extracted from IPTs, the next hops for further interaction between IPT operators and potential customers (victims).  Our analysis focuses on answering two questions. One is what security risks these next hops will expose a victim to. The other is what evasion techniques have been adopted in these next hops to avoid detection or blocking. 

\subsection{IPT Websites}
Also, as detailed in \S\ref{subsec:method_cpt_analyzer}, a website infiltrator was scheduled weekly to dynamically visit each IPT website using a headless browser and save the results as a snapshot which consists of a screenshot of the landing page as well as the traffic logs. Among all the 16,335 website domains embedded in IPTs, we identified 172 different top-level domains (TLDs), with the top five most frequent occurrences being com (60.60\%), cc (7.10\%), xyz (4.13\%), net (3.36\%), and vip (2.20\%).
We scrutinized 9,149 websites collected in 2022, of which 5,721 could be successfully resolved to IP addresses. Subsequently, 125,715 snapshots were successfully captured for 5,055 out of all 5,721 IPT websites, while the remaining websites were found to be unreachable during our infiltration period.

\finding{IPT websites are equipped with various evasion techniques and can redirect victims to diverse categories of unsolicited or illicit content or services.}

\begin{table}
    \centering
    \footnotesize
    \caption{The categories of IPT websites.}
    \label{tab:cybercrime_website_distribution}
    \begin{threeparttable}
        \begin{tabular}{cr|cr}
            \toprule
            Category & \% Websites &  Category & \% Websites \\
            \midrule
            Gambling & 22\% & Financial Fraud & 2\%\\
            Access Blocked~\tnote{1} & 21\% & Drug Sales &  1\%\\
            Benign &  18\% &  Fake Account & 1\%\\
            Domain Expired & 14\% & Hacking Service & 0.75\% \\
            Sex~\tnote{2}& 11\% & Advertisement & 0.5\%\\
            Redirection Page & 7\% & Data Theft &  0.33\% \\
            Black Hat SEO & 2\% & Ghost Writing  & 0.17\%\\
            \bottomrule
        \end{tabular}
        \begin{tablenotes}
            \item [1] This is the group of websites that have blocked our visit.
            \item [2] Websites in this group either host pornography content or offer sex services.
        \end{tablenotes}
    \end{threeparttable}
\end{table}

\subject{Security risks of IPT websites.} We profile the security risks of IPT websites through categorizing their web content. We thus randomly sampled 1,200 IPT websites, manually looked into their snapshots, and categorized them based on their web content. The category distribution of these sampled websites is listed in Table~\ref{tab:cybercrime_website_distribution}.  In total, among the top 14 categories with most IPT websites, 8  are either illicit or unsolicited, e.g., gambling, sex, drug sales, hacking service, etc. Such an observation is aligned with that observed for IPTs (\S\ref{subsec:landscape}). 
Particularly,  gambling websites have the largest share of 22\%. Also, they are widely distributed in natural languages and target customers of different countries including China, Vietnam, Thailand, Korea, etc. What is followed are IPT websites either promoting sex services or hosting pornography content, which accounts for 11\%.  For instance, website \textit{ncao3.com} is used to host child pornography content, which is promoted in 1,428 different IPTs.  On the other hand, \textit{topjqk.com} is a website advertising hacking services, while \textit{a9play2u.com} is a gambling website.

One thing to note, 21\% websites blocked our automatic access, as revealed by either the HTTP response code or the error messages rendered on the landing page. 
Also,  snapshots of another 18\% websites appear to be benign. We then manually visited a sampled set of such websites and found out that the ultimate landing page for many of them is not benign. Instead, we have found many of IPT websites have deployed various evasion techniques which render our automatic visits either blocked or redirected to a deceptively benign web page. More details will be discussed as below. 
In addition, 7\% websites land on a seemingly benign page that contains only multiple redirection links, and clicking these links will redirect the visitor to the final webpage for illicit services or goods. We categorize such pages as redirection pages. 
Fig.~\ref{fig:redirection_cases} presents an example of redirection pages along with the ultimate landing page after clicking the redirection links. 
Since the redirection pages appear to be benign, without visiting the redirection links, there is no way to decide whether the IPT website is illicit or not.

\begin{figure}[!htbp]
    \centering
    \subfigure[
        The redirection page of visiting IPT website \textit{dk8.vn}.
        \label{fig:redirection_case_2_1}
    ]{
        \includegraphics[width=.45
        \columnwidth]{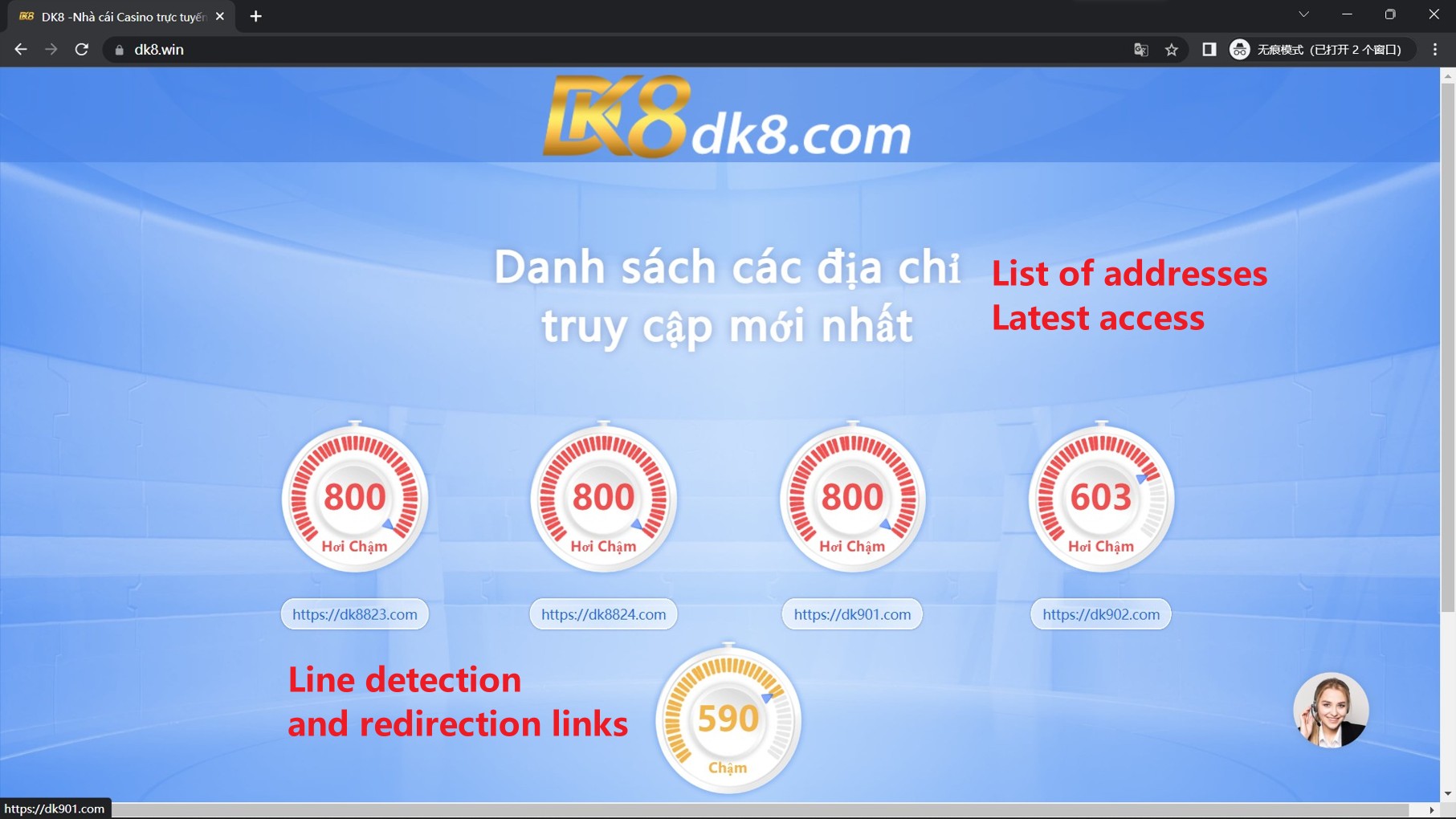}
    }
    \hfill
    \subfigure[
        The final landing page of the redirection link in Figure~\ref{fig:redirection_case_2_1}.
        \label{fig:redirection_case_2_2}
    ]{
        \includegraphics[width=.45
        \columnwidth]{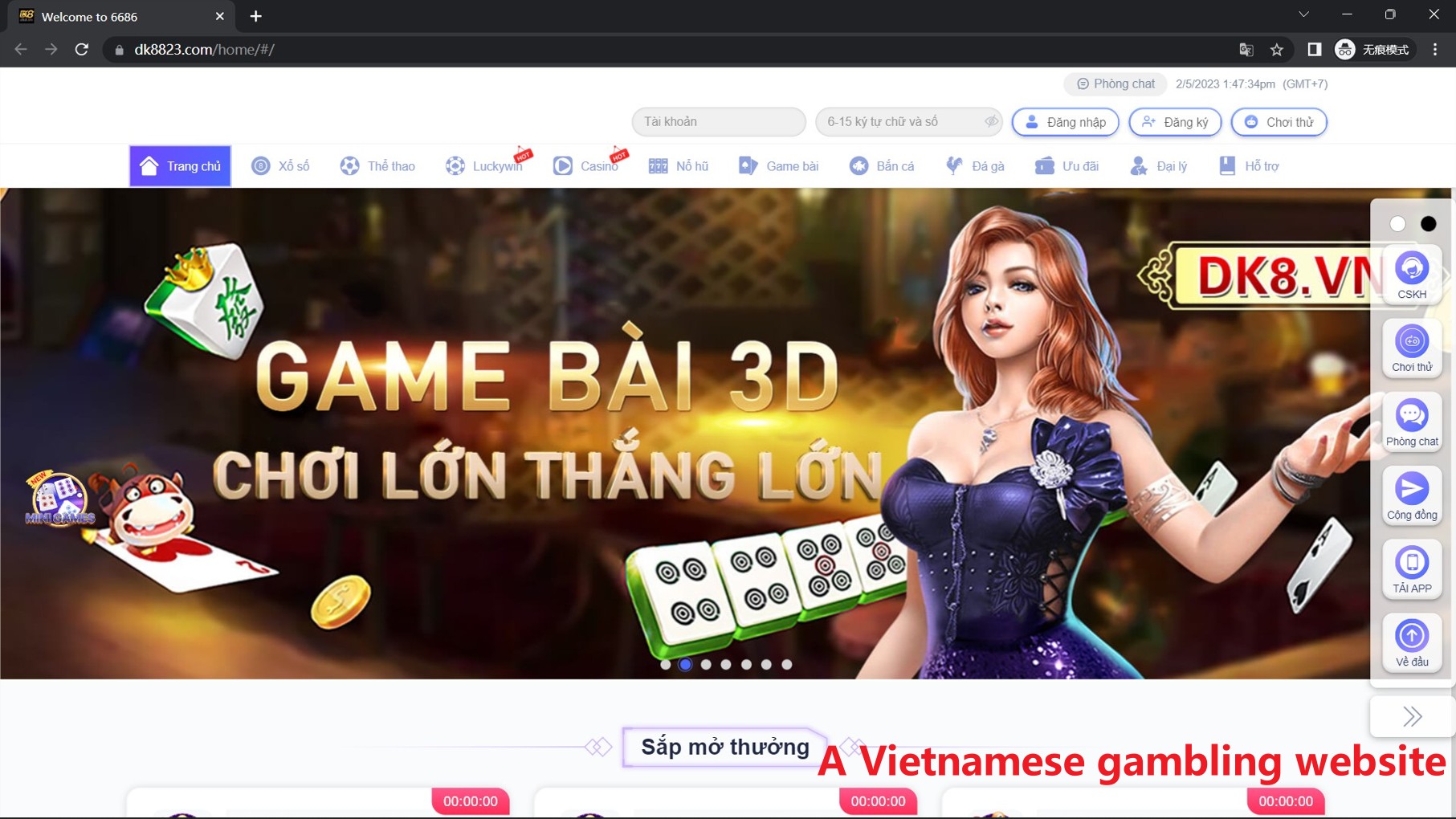}
    }
    \caption{A Case of redirection pages.}
    \label{fig:redirection_cases}
\end{figure}

\subsubject{Malicious mobile apps distributed through IPT websites.} As elaborated in \S\ref{subsec:method_cpt_analyzer}, when the landing page of an IPT website is found to promote a mobile app, the Android version of the respective app will be manually downloaded and further analyzed.
In total, we have manually collected 200 distinct Android APK files from 1,200 different IPT websites. Among these APK files, 123 are gambling apps and 66 are pornography apps.

Subsequently, we queried VirusTotal with the hash values of these APKs and found out that only 68 had historically been observed by VirusTotal. We then uploaded the payloads of the missing 132 APKs to VirusTotal for further analysis.
Among all these APKs, 98 were detected as malicious with seven types of threats.
The most prevalent threat observed is the trojan type, accounting for 32.38\%, which typically disguises itself as a legitimate app, and once the user opens this app, the malware is activated.
The second most frequent threat type is riskware (24.76\%)
While these programs are not inherently designed to be malicious, they do have functions that can be exploited for malicious purposes. So it can be further categorized. 
Fifteen out of the total 52 APK files under riskware are flagged as using untrustworthy hardening tools 
(such as 360-Jiagu~\footnote{https://jiagu.360.cn/}), 
which are commonly used to protect APK files from being reverse-engineered and modified. However, these tools can also be exploited by malicious actors to hide or obfuscate malicious code. 
Among APKs flagged as riskware, 12 are further flagged as \textit{riskware-scamapp}, indicating that they may be used for online fraudulent activities. 
The full categories of these threats, along with the number of alerted APKs in each category, are listed in Appendix~\ref{appendix:mobile_threat}.

\subject{Evasion techniques of IPT websites.} We have observed the adoption of multiple evasion techniques by IPT websites, likely in an attempt to escape from detection radars. And representative evasion techniques include the lengthy redirection chain, iframe cloaking, and location-based access control. Below, we provide more details on these evasion techniques along with demonstrative examples.

\subsubject{The lengthy redirect chain.} Different from typical benign websites, many IPT websites have a very long redirect chain before reaching the landing page. Table~\ref{tab:website_redirection_dist} presents the distribution of 125,715 website snapshots across the length of the redirect chain. Here, the redirect length is defined as the number of distinct FQDNs observed when visiting a given  website, and the FQDN of the landing page will be counted if it is different from the FQDN of the given IPT website.  Almost 13.62\% snapshots involve three or more redirection hops, while 32 snapshots have even 10 or more redirection hops. 
For instance, when a visitor tries to access the website \textit{go.win}, they will be redirected through 8 different FQDNs before finally arriving at the landing page of the gambling website \textit{kingpro.vip}.
One possible purpose for implementing a lengthy redirect chain is to defer automatic crawling attempts and therefore evade potential content moderation. For instance, Google only follows up to 5 hops in a redirect chain for every crawling attempt~\cite{google_redirect_hops}. Websites with 6 or more redirection hops in a row can likely escape Google's detection radar. 

In addition to the lengthy redirect chain, we also observed that multiple different IPT websites could eventually jump to the same landing page. For instance,  88 IPT websites have been redirected to \textit{https://linktr.ee/ak5537}, a webpage advertising a black hat SEO service.
Also, many IPTs embedding these 88 websites are of categories rather than black hat SEO, e.g., gambling and financial fraud. Another case is \textit{www.ab0030.com}, a landing page about online gambling,  which is redirected from 44 IPT websites. However, different from the first example, the parental IPTs of these original websites promote the content of the same category as the landing page. 
Further investigation makes us believe that RSP services have promoted in advance many IPTs of diverse categories, even before customers have placed any website promotion orders. Also, IPTs of the same category will likely have a common set of websites $W_{SEO\text{, }C}$ embedded wherein $C$ denotes a category. Also, all these websites are under the direct control of the RSP operator. Then, upon successful RSP attacks, search users visiting $W_{SEO{, }C}$ can be redirected to any website by following the configuration of the RSP operator.
For instance, by default, such a visit can be directed to the landing page promoting the RSP service itself, which is the case in the first example. Instead, when a customer came to request a promotion for a gambling website, depending on how much the customer has paid, traffic towards $W_{SEO{, }gambling}$ could be partially redirected to the to-be-promoted actual gambling website. In a nutshell, this game consists of two steps. The first step is to poison search engines and harvest visiting traffic of search users, and the second step is to monetize visiting traffic by redirecting it to customers' websites upon request.
This service model also explains another observation that an IPT website can be redirected to different landing pages when visiting it across time. Two landing pages are considered as different when they have distinct FQDNs.
In total, during our months of infiltration, 949(18.77\%) IPT websites have two or more distinct landing pages observed, while 110(2.18\%) have five or more.

\begin{table}
    \centering
    \footnotesize
    \caption{The distribution of IPT websites across the length of their redirect chains.}
    \label{tab:website_redirection_dist}
    \resizebox{\linewidth}{!}{
    \begin{threeparttable}
        \begin{tabular}{crrrr}
            \toprule
            Redirects & Snapshots & \% Snapshots & Websites & \% Websites\\
            \midrule
            $\ge 10$ & 31 & 0.02\% & 2 & 0.04\%\\
            $\ge 5$ & 1,144 & 0.91\% & 65 & 1.29\%\\
            $\ge 4$ & 3,615 & 2.88\% & 210 & 4.15\%\\
            $\ge 3$ & 17,118 & 13.62\% & 901 & 17.82\%\\
            $\ge 2$ & 61,926 & 49.26\% & 3,056 & 60.45\%\\
            $\ge 1$ & 125,715 & 100.00\% & 5,055 & 100.00\%\\
            \bottomrule
        \end{tabular}
    \end{threeparttable}
    }
\end{table}

\subsubject{Iframe cloaking.}
Some websites utilize the HTML iframe tag to embed illicit webpages into benign pages. Fig.~\ref{fig:website_evasion} illustrates a typical example where the website \textit{http://nbjsfl.com/} initially appears benign, but later an iframe is rendered with gambling content, and the original content is replaced without any page jump. If a URL detection system stopped rendering the iframe when visiting this website, this website would likely be considered benign.

\subsubject{Location-based access control.} Some websites limit access based on the visitor's IP location. For instance,  when attempting to access \textit{https://kkh1.xyz/} from the United States, the final landing page is \textit{google.com}. However, when accessing the site from China, users are redirected to a page containing multiple redirect links towards child pornography websites.  We then compared the results of our dynamic website crawlers deployed on a US server and a server located in China on Apr. 4, 2023. Out of 4,827 IPT websites reachable on that day, 546 (11.31\%) have very different web content. 

\begin{figure}
    \centering
    \includegraphics[width=.9\columnwidth]{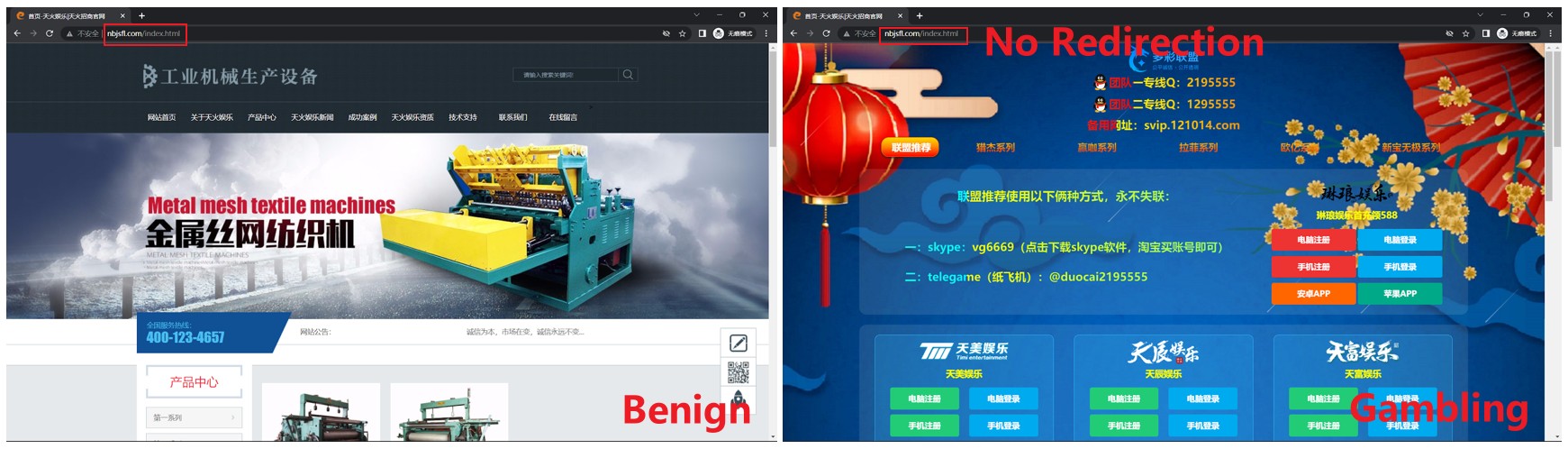}
    \caption{Iframe cloaking.}
    \label{fig:website_evasion}
\end{figure}



\subsection{Instant Messaging Accounts}
Similar to IPT websites, IM accounts embedded in IPTs can also expose victims to illicit goods and services of diverse categories. On the other hand, a large volume of illicit promotion activities has also been observed on Telegram. 

\subject{WeChat and QQ accounts}. In total, our IPT contact extractor has identified 23,632 WeChat IDs and 1,552 QQ IDs. However, due to the unavailability of programmable APIs for either platform,  we conducted manual case studies. 

\begin{figure}
    \centering
    \subfigure[A WeChat account related to forging diplomas.]{
    \label{fig:wechat}
    \includegraphics[width=.4\columnwidth,height=.16\textwidth]{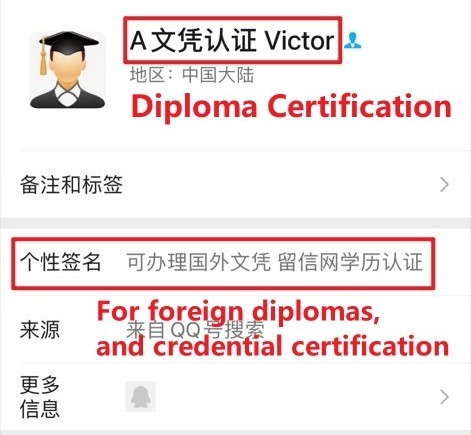}
    }
    \subfigure[A QQ account related to forging diplomas.]{
    \label{fig:qq}
    \includegraphics[width=.4\columnwidth,height=.2\textwidth]{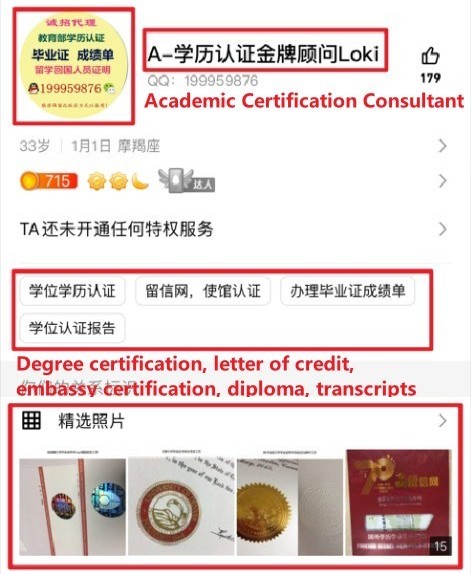}
    }
    \caption{Examples of WeChat and QQ accounts.}
    \label{fig:wechat_and_qq}
\end{figure}

We manually looked into the public profiles of 250 randomly selected WeChat accounts, among which 64 had already been suspended by the time of our visit. Among the 186 left ones, 154 can be identified with high confidence as accounts promoting illicit services or goods. As shown in Fig.~\ref{fig:wechat}, the profile of a WeChat account can reveal much information regarding the respective goods or services under provision. Then, when ordered by the number of WeChat accounts, the top categories of illicit goods or services are Fake Certificate (33\%), Surrogacy (17\%), Counterfeit Goods (11\%), and Sex Service (10\%).

For QQ, we examined randomly selected 200 accounts, and 6 were suspended by the time of our visit. Among the left ones, 172 (89\%) have been confirmed with high confidence to be associated with illicit services and goods, by the consideration of the username, avatar, user tag, and the photo wall in their profiles, as demonstrated in the Fig.~\ref{fig:qq}. And the top categories of illicit goods and services include Fake Certificate (40\%), Sex Service (16\%), and Gambling (12\%).

\subject{Telegram accounts.} As detailed in \S\ref{subsec:method_cpt_infiltrator}, leveraging Telegram's programmable APIs, we are allowed to automatically infiltrate the 
 Telegram accounts embedded in IPTs. 
 In total, 4,732 Telegram accounts have been infiltrated,  which consist of 1,507 users, 231 bots, 2,333 channels, and 661 groups. For each account, the profile is retrieved, and so are historical messages for group/channel accounts, which leads to the collection of over 14 million historical messages for the time period between January, 2022 and  March, 2023. These historical messages along with the account profiles allow us to understand what security risks a victim of IPTs may be further exposed to by contacting these embedded Telegram accounts. 
Within these messages, we identified 197,689 distinct URLs, most of which were links promoting other Telegram accounts. Therefore, we then sampled 8,472 (4.29\%) distinct URLs and continued to crawl them, which identified a total of 2,322 new Telegram accounts, including 375 users, 25 bots, 1,026 channels, and 896 groups. 

Given the large volume of Telegram messages, we then utilized the multi-label IPT classifier introduced in Section~\ref{subsec:method_cpt_analyzer} to classify these messages. 
As listed in Table~\ref{tab:tg_cybercrime_messages},  over 6 million (43.75\%) messages are mapped to categories of illicit goods and services, especially money laundering (31.96\%), black hat SEO (17.57\%), data theft (13.93\%), gambling (12.46\%), and financial fraud (8.49\%). Besides, the respective Telegram channels have over 29 million subscribers while there are 600K distinct members in the respective Telegram groups. All these data points suggest that the Telegram platform is extensively used by miscreants for not only the promotion of their illicit goods and services but also the communication with their customers. 

\begin{table}
    \centering
    \caption{The distribution of Telegram group/channel messages over the categories of goods and services. 
    }
    \label{tab:tg_cybercrime_messages}
  \resizebox{\linewidth}{!}{
    
    \begin{threeparttable}
        \begin{tabular}{cr|cr}
            \toprule
            Category & \% Telegram Msgs & Category & \% Telegram Msgs  \\
            \midrule
            Money Laundering & 31.96\% & Sex Service & 4.87\%\\ 
            Black Hat SEO \& Advertisement & 17.57\% & Hacking Service & 3.57\%\\
            Data Theft & 13.93\% & Fake Account & 3.53\%\\
            Gambling & 12.46\% & Others & 1.78\%\\
            Financial Fraud & 8.49\% & Counterfeit Goods & 0.70\%\\
            \bottomrule
        \end{tabular}
    \end{threeparttable}
    }
\end{table}

\section{Discussion}
\label{sec:discuss}
\subject{Mitigation recommendations.}
For search engine operators, more efforts should be invested in filtering out IPTs and preventing them from reaching regular search users. In this case, the series of our binary IPT classifiers can contribute to this task. Particularly, when detection efficiency is the bottleneck, the feature-based Random Forest classifier can be deployed to enable high-throughput IPT detection. On the other hand, when efficiency is not a matter but the accuracy, the BERT-based classifier can help achieve higher detection performance. Also, for a website with a legitimate URL reflection scheme, simple measures may also be taken to avoid abuse from  miscreants. For instance, the text value of a reflection parameter should be omitted when rendering the resulting webpage, as long as it is abnormal, e.g., it has no search results in the scenario of on-site searching, or it is not a valid hashtag in the scenario of hashtag searching. Once deployed with such a measure, the IPTs (i.e., the search keywords) will not be rendered as  part of the webpage and will not be indexed by the search engines.

\subject{Limitations and future works}. Our study still suffers from some limitations. \arevised{Firstly, some forum sites, such as Medium, incorporate the title of a blog post into its URL. Consequently, for such forum sites,  the URL of a forum spamming post will look like reflected search poisoning. As our IPT hunter relies solely on the URL parameters to detect IPTs,  it may not distinguish an RSP case from a forum spamming post that has its title embedded in the URL. Upon recognizing this issue, we conducted a manual inspection of a subset of the RSP dataset and discovered that approximately 5\% belong to such kinds of forum spamming, while the left majority are true RSPs. We thus believe this issue has a minimal impact on our measurements.
} 
Secondly, just like any machine learning based system, our machine learning classifiers may be vulnerable to various adversarial attacks, e.g., adversarial examples, and poisoning attacks. For instance, IPT operators may adjust how they compose IPTs to make our manually crafted feature set less effective and thus evade our IPT hunter. We leave it as a future work to fully evaluate the adversary resistance of our machine learning models.
Furthermore, many IPTs were found to have been promoted simultaneously through other channels besides search engines, e.g., online social networks, for which, extra data can be found in Appendix~\ref{appendix:other_channel}. And it will be interesting to further explore how IPTs distribute through other channels and how they can evade the content moderation of respective channels (e.g., online social networks). 

\subject{Responsible disclosure.} We have conducted responsible disclosure to two groups of relevant parties. One is the communication with the 4 search engines that are poisoned by IPTs. By this writing, Bing has responded to our disclosure and stated that it is working to address this poisoning issue, which may explain the obvious drop in IPTs observed for Bing between our first-phase hunting in November, 2022 and the second-phase hunting in November, 2023.  However, for the other three search engines, we have yet to receive any concrete response.  
Then, it is the disclosure towards the instant messaging platforms on which a large number of IPT accounts are observed, especially WeChat, QQ, and Telegram. 
Despite receiving some preliminary replies, the disclosure towards these instant messaging platforms is ongoing, and we will update it here once the final concrete results are available.


\subject{Data and code release.} To prevent miscreants from launching new RSP attacks, we hide the vulnerable URSes and make the rest of the data public which include the IPT search keywords, the IPTs, the IPT contacts of various categories and the messages collected from Telegram. Also, in our study, multiple machine learning models have been built to detect and understand IPTs, e.g., the binary/multi-label IPT classifiers, and the IPT contact extractor. For these models, we open-source the scripts for training and testing  as well as the ground-truth datasets. The data and code are available at \url{https://github.com/ChaseSecurity/Reflected-Search-Poisoning-for-Illicit-Promotion}.
\section{Conclusion}
\label{sec:conclusion}
Through this study, we can conclude that reflected search poisoning (RSP) has been extensively exploited to free ride high-rank websites and distribute illicit promotion texts (IPTs) that are large-scale, available across search engines, as well as being diverse in categories of goods and services. Also, most services and goods promoted in IPTs belong to categories that are illicit or illegal, while regular search engine users can be exposed to such IPTs at a concerning extent.  Also, victims of IPTs can be further exposed to harmful content and illegal services when following the next-hop contacts embedded in IPTs.  All these results highlight the necessity of more efforts to be invested in the fight against illicit promotion and the underground economy.

    \bibliographystyle{IEEEtranS}
    \bibliography{ref}
    \appendix
\subsection{Extra Details for the Binary IPT Classifier}
\subsubsection{Manually Crafted Features for the Binary IPT Classifier}
\label{appendix:IPT_features}

\subject{Length in characters}. 
Most IPTs are lengthy with a median length of 46 characters, but benign texts are more likely to be shorter. 

\subject{Number of bracket-like characters} IPTs are more likely to have several bracket-like characters  (e.g. \{, \},  [, ], 【, 】,『, 』) to highlight contact entities. 

\subject{Number of URLs in IPT.}
Some IPTs embedding website contact entities may have URLs.


\subject{Number of numeric characters.}
IPTs with wechat, qq or telephone contacts often have numeric characters. 

\subject{Number of emojis and unicode symbols.}
Some IPTs use emojis or meaningless symbols to make themselves eye-catching.

\subject{Number of patterns about next-hop instant messaging accounts.} These patters include ['微信', 'q微', '扣微', '微', '薇', '扣扣', 'qq', 'www', 'com', 'fun', 'cc', 'tg', 'telegram', '飞机', '@', '网', 'V信'].

\subject{If there is a file suffix.} Common file suffixes for URLs include ['.html', '.shtml', '.htm', '.php', '.pdf', '.jpg', '.jpeg', '.png', '.xlsx', '.docx', '.pptx', '.xml']. Rather than benign URLs, IPTs tend to have no such kinds of suffixes. 

\subsubsection{Hyperparameters for the Binary IPT Classifier}
\label{appendix:hyperparams_binary_ipt}
\ 
\newline
Below, we summarize the hyperparameters for training the series of binary IPT classifiers.

\subject{Decision Tree}
\begin{itemize}[leftmargin=*]
    \item criterion: gini
    \item splitter: best
    \item max\_depth: None
    \item min\_samples\_split: 2
    \item min\_samples\_leaf: 1
\end{itemize}

\subject{Random Forest}
\begin{itemize}[leftmargin=*]
    \item n\_estimators: 91
    \item criterion: gini
    \item max\_features: 1
    \item max\_depth: None
    \item min\_samples\_leaf: 1
    \item min\_samples\_split: 2
    \item class\_weight: balanced
\end{itemize}

\subject{AdaBoost}
\begin{itemize}[leftmargin=*]
    \item best\_estimator: None
    \item n\_estimators: 100
    \item learning\_rate: 1.0
    \item algorithm: SAMME.R
\end{itemize}

\subject{SVM}
\begin{itemize}[leftmargin=*]
    \item penalty: l2
    \item loss: squared\_hinge
    \item dual: True 
    \item tol: 0.0001
    \item C: 1.0
    \item multi\_class: ovr
    \item fit\_intercept: True
    \item max\_iter: 10000
\end{itemize}

\subject{Fine tuning the BERT model}
\begin{itemize}[leftmargin=*]
    \item model\_name: bert-base-multilingual-cased 
    \item vocab\_size: 30522
    \item hidden\_size: 768
    \item num\_hidden\_layers: 12
    \item num\_attention\_heads: 12
    \item intermediate\_size: 3072
    \item hidden\_act: gelu
    \item hidden\_dropout\_prob: 0.1
    \item attention\_probs\_dropout\_prob: 0.1
    \item training\_epoch: 50
    \item learning\_rate: 4e-5
    \item training\_batch\_size: 8
\end{itemize}

\subsection{The List of Features for the Contact Segment Classifier}
\label{appendix:contact_features}
\subject{Length in characters}. 
The contact segments are more likely to be shorter than other segments of IPTs.

\subject{Number of URLs in IPT.}
Some segments of IPTs embedding website contact entities may have URLs.

\subject{Number of non-alphanumeric characters.}
The contact segments are more likely to be less non-alphanumeric characters than other segments of IPTs.

\subject{Number of alphanumeric characters.}
The contact segments are more likely to be more alphanumeric characters than other segments of IPTs.

\subject{Number of numeric characters.}
Segments with wechat, qq or telephone contacts often have numeric characters. 

\subject{Number of contact indicators.}
The indicators of next-hop contacts include ['微信', 'q微', '扣微', '微', '薇', '扣扣', 'qq', 'com', 'fun', 'cc', 'hash', 'tg', 'telegram',  '飞机', '@', ‘网’, ‘复制’]. Those marks often represent the contact information category, e.g., \textit{tg} for Telegram, and '扣扣' for QQ.

\subject{Number of some common punctuation marks in terms.}
 Various punctuation symbols are used to separate an IM mark and the respective IM account identifier, e.g., the colon in "询telegram:@ts775". Symbols under our consideration include [., :, ：, ·, ͺ, -].

\subject{If there is a file suffix.} Common file suffixes for URLs include ['.html', '.shtml', '.htm', '.php', '.pdf', '.jpg', '.jpeg', '.png', '.xlsx', '.docx', '.pptx', '.xml']. Rather than benign URLs, IPTs tend to have no such kinds of suffixes. 

\subsection{Category-Specific Performance for the Multi-Label IPT Classifier}
\label{appendix:performance_cybercrime_classifier}
Table~\ref{tab:performance_cybercrime_classifier} shows the per-class performance of the multi-label IPT classifier. It reports precision, recall, and F1-Score for different categories of services and goods.

\begin{table}
    \caption{The per-class performance for the multi-label IPT classifier.}
    \label{tab:performance_cybercrime_classifier}
  \resizebox{\linewidth}{!}{
    
    \centering
    \begin{threeparttable}
        \begin{tabular}{cccc}
            \toprule
            Category & Precision & Recall & F1-Score\\
            \midrule
            Benign & 0.95 & 0.96 & 0.96\\
            Black Hat SEO \& Advertisement & 0.95 & 0.95 & 0.95\\
            Counterfeit Goods & 0.83 & 0.88 & 0.86\\
            Data Theft & 0.94 & 0.96 & 0.95\\
            Fake Account & 1.00 & 0.92 & 0.96\\
            Fake Certificate & 0.98 & 0.91 & 0.94\\
            Financial Fraud & 0.70 & 0.78 & 0.74\\
            Gambling & 0.94 & 0.95 & 0.95\\
            Hacking Service & 0.90 & 0.85 & 0.88\\
            Drug Sales & 0.97 & 0.83 & 0.89\\
            Sex Service & 0.94 & 0.93 & 0.94\\
            Surrogacy Service & 0.94 & 0.94 & 0.94\\
            Weapon Sales & 1.00 & 0.83 & 0.91\\
            Money Laundering & 0.91 & 0.91 & 0.91\\
            Others & 0.97 & 0.97 & 0.97\\
            \bottomrule
        \end{tabular}
    \end{threeparttable}
    }
\end{table}

\subsection{Products Promoted in IPTs}
\label{appendix:IPT_products}

\subject{Black Hat SEO \& Advertisement.} Services in the category of the Black Hat SEO \& Advertisement support illicit promotion across many social media platforms which include:
\begin{itemize}\setlength{\itemsep}{0pt}\setlength{\parsep}{0pt}\setlength{\parskip}{0pt}
    \item Facebook
    \item Telegram
    \item WhatsApp
    \item Zalo
    \item Skype
    \item YouTube
    \item Google
    \item Signal
    \item Twitter
    \item Shopee
    \item Instagram
    \item TikTok
    \item QQ
\end{itemize}

\subject{Fake Certificate.} The list of products as observed through manual study include:


\begin{itemize}\setlength{\itemsep}{0pt}\setlength{\parsep}{0pt}\setlength{\parskip}{0pt}
    \item The Resident Identity Card in China
    \item Visa
    \item Passport
    \item Birth Certificate
    \item Title Deed
    \item Household Registration
    \item Letter of Employment
    \item Police Clearance Certificate
    \item Resignation Letter
    \item Professional Qualification Certificate
    \item Invoice
    \item Marriage Certificate
    \item Divorce Certificate
    \item Driver's License
    \item Business License
    \item Diploma
    \item Transcript
    \item Tuition Bill
    \item Medical Record
\end{itemize}

\subject{Fake Account.} The list of products as observed through manual study include:
\begin{itemize}\setlength{\itemsep}{0pt}\setlength{\parsep}{0pt}\setlength{\parskip}{0pt}
    \item SIM Card
    \item Bank Card
    \item Amazon Account
    \item Google Account
    \item Twitter Account
    \item Telegram Account
    \item LinkedIn Account
    \item Alipay Account
    \item QQ Account
    \item WeChat Account
    \item TikTok Account
    \item KuaiShou Account
    \item Xiaohongshu Account
    \item Uber Account
    \item CMB Account
    \item Snapchat Account
    \item Skype Account
    \item OkCupid Account
    \item Momo Account
    \item PayPal Account
\end{itemize}
    


\subject{Drug Sales.} The list of products as observed through manual study include:
\begin{itemize}\setlength{\itemsep}{0pt}\setlength{\parsep}{0pt}\setlength{\parskip}{0pt}
    \item Marijuana
    \item Methamphetamine
    \item Sedatives
    \item Gamma-hydroxybutyrate (GHB)
    \item Sevoflurane
    \item Alprazolam
    \item Midazolam
    \item Desflurane
    \item Triazolam
    \item Clonazepam
    \item Viagra
    \item Thuocduame
    \item Estazolam
\end{itemize}

\subject{Weapon Sales.} The list of products involved as observed through manual study include:
\begin{itemize}\setlength{\itemsep}{0pt}\setlength{\parsep}{0pt}\setlength{\parskip}{0pt}
    \item Gun
    \item Bullet
    \item Slingshot
    \item Electric Baton
    \item High-pressure Gas Cylinder Valve
\end{itemize}

\subsection{The Differences among Search Engines in IPTs}
\label{appendix:cross_SE_IPT_distribution}
Fig.~\ref{fig:cross_SE_heatmap} shows the IPT overlap rate cross different search engines. In the heatmap, each cell denotes the intersection ratio between a pair of two search engines with the denominator being the search engine of the respective row.  And we can see that most IPTs observed on a search engine are exclusive to IPTs of other search engines, which is likely due to the different strategies taken by search engines in terms of indexing and filtering webpages. 
\begin{figure}[H]
    \centering
    \includegraphics[width=.9\columnwidth]{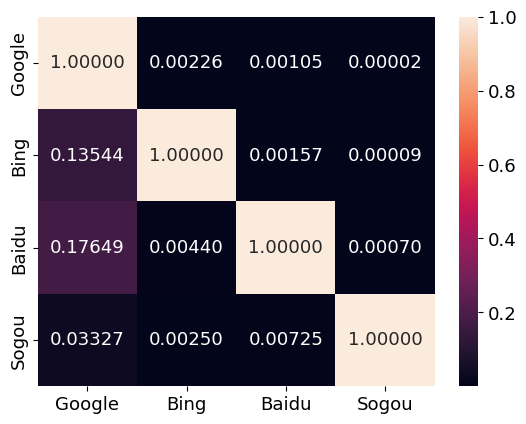}
    \caption{The IPT overlap rates for pairs of search engines.}
    \label{fig:cross_SE_heatmap}
\end{figure}

\subsection{The Category Distribution for IPTs in Different Languages}
\label{appendix:category_distribution}
Table~\ref{tab:language_category} lists  the category distribution of IPTs in different natural languages. The most promoted category in Chinese is Gambling, accounting for 26.66\% of observed IPTs. The second and third most promoted categories in Chinese IPTs are Fake Certificate (26.29\%) and Sex Services (20.92\%), respectively. And in Korean, the most promoted category is Sex Services, accounting for 71.95\% of observed IPTs. This is because in Korea, the underground industry of phone sex services is relatively large in scale, but other types of grey industries are relatively small in scale. Furthermore, categories of Gambling (52.98\%) and Financial Fraud (27.01\%) account for a very high percentage of English IPTs.
\begin{table}
    \footnotesize
    \centering
    \caption{Top 10 most promoted categories for IPTs in different natural languages }
    \label{tab:language_category}
    \resizebox{\linewidth}{!}{
    \begin{threeparttable}    
    \begin{tabular}{crcrcr}
      \toprule
     \multicolumn{2}{c}{Chinese} & \multicolumn{2}{c}{Korean} & \multicolumn{2}{c}{English}  \\
      \midrule
        Gambling & 26.66\% &  Sex & 71.95\% & Gambling & 52.98\%\\
         Fake Certificate & 26.29\% & SEO \& Ads & 8.83\% & Financial Fraud & 27.01\%\\
        Sex & 20.92\% & Drug Sales & 6.60\% &  Drug Sales & 8.21\%\\
         SEO \& Ads~\tnote{1} & 9.93\% & Financial Fraud & 6.40\% & Sex & 5.01\%\\
        Fake Account & 4.97\% & Gambling & 3.52\% & Counterfeit& 2.26\%\\
         Hacking Service & 3.07\% & Counterfeit & 1.24\% & Fake Account & 1.29\%\\
         Data Theft & 1.71\% & Fake Account & 0.73\% & Others & 1.03\%\\
        Counterfeit& 1.62\% & Hacking Service & 0.40\% & Fake Certificate & 0.90\%\\
         Surrogacy & 1.57\% & Others & 0.22\% & Money Laundering & 0.57\%\\
         Others & 1.08\% & Fake Certificate & 0.03\% & SEO \& Ads & 0.91\%\\
       \bottomrule
    \end{tabular}
        \begin{tablenotes}
            \item [1] It denotes Black Hat SEO \& Advertisements.
            \item [2] Counterfeit Goods.
        \end{tablenotes}
    \end{threeparttable}
}
\end{table}

\subsection{More Details for Mostly Abused Websites}
\label{appendix:top_abused_website}
Table~\ref{tab:most_abused_urls} lists top 10 mostly abused URL reflection schemes, while Table~\ref{tab:most_abused_domains} gives top 10 mostly abused apex domains.

\begin{table}[H]
    \centering
    \footnotesize
    \caption{Top 10 mostly abused URL reflection schemes.}
    \label{tab:most_abused_urls}
        \begin{tabular}{ccc}
        \toprule
        URL Scheme & \% IPTs & \% RSPs \\
        \midrule
        https://search.bilibili.com/all?keyword=XX & 0.5002\% & 0.4629\%\\
        https://spankbang.com/s/XX & 0.4522\% & 0.4180\%\\
        https://www.goodreads.com/quotes/tag/XX & 0.4330\% & 0.3866\%\\
        https://us.puma.com/us/en/search?q=XX & 0.3856\% & 0.3527\%\\
        https://pngtree.com/free-png-vectors/XX & 0.3793\% & 0.3387\%\\
        https://www.cnrtl.fr/definition/XX & 0.3553\% & 0.3173\%\\
        https://www.royal.uk/search?tags[]=XX & 0.3532\% & 0.3615\%\\
        https://olx.ba/pretraga?trazilica=XX & 0.3397\% & 0.3043\%\\
        https://www.thedp.com/page/search?q=XX & 0.3319\% & 0.2967\%\\
        https://www.pixiv.net/tags/XX & 0.3278\% & 0.2988\%\\
        \bottomrule
      \end{tabular}
\end{table}

\begin{table}[H]
    \centering
    \footnotesize
    \caption{Top 10 mostly abused apex domains.} 
    \label{tab:most_abused_domains}
    \begin{tabular}{ccc}
        \toprule
        Domain & \% IPTs & \% RSPs\\
        \midrule
        baidu.com & 1.0757\% & 1.3262\%\\
        pixnet.net & 1.0594\% & 0.9512\%\\
        gfycat.com & 0.6437\% & 0.6406\%\\
        facebook.com & 0.5838\% & 0.5516\%\\
        pixiv.net & 0.5790\% & 0.5581\%\\
        youtube.com & 0.5729\% & 0.5132\%\\
        cnrtl.fr & 0.5613\% & 0.5014\%\\
        bilibili.com & 0.5500\% & 0.5125\%\\
        spankbang.com & 0.4842\% & 0.4586\%\\
        goodreads.com & 0.4330\% & 0.3866\%\\
        \bottomrule
    \end{tabular}
\end{table}

\subsection{Extra Data for Mobile Threats}
\label{appendix:mobile_threat}

Table~\ref{tab:mobile_threat} presents a comprehensive list of mobile threats associated with suspicious Android APKs, including the number and percentage of APKs affected by each type of threat. It comprises seven rows, each representing a distinct mobile threat. The most widespread threat detected in this study is the trojan, followed by riskware, malware, and evader. Adware, malformed, and spyware are the least frequent types of mobile threats encountered in the APKs. 

\begin{table}[H]
    \centering
    \footnotesize
    \caption{The list of mobile threats alarmed by VirusTotal for Android apps distributed through IPT websites.}
    \label{tab:mobile_threat}
    \begin{threeparttable}
        \begin{tabular}{crr}
            \toprule
            Mobile Threat & APKs & \% APKs\\
            \midrule
            Trojan & 68 & 32\%\\
            Riskware & 52 & 25\%\\
            Malware & 40 & 19\%\\
            Evader & 26 & 12\%\\
            Adware & 12 & 6\%\\
            Malformed & 8 & 4\%\\
            Spyware & 4 & 2\%\\
            \bottomrule
        \end{tabular}
    \end{threeparttable}
\end{table}

\begin{figure}[H]
    \centering
    \subfigure[Weibo.]{
    \label{fig:other_promotion_weibo}
    \includegraphics[width=.35\columnwidth]{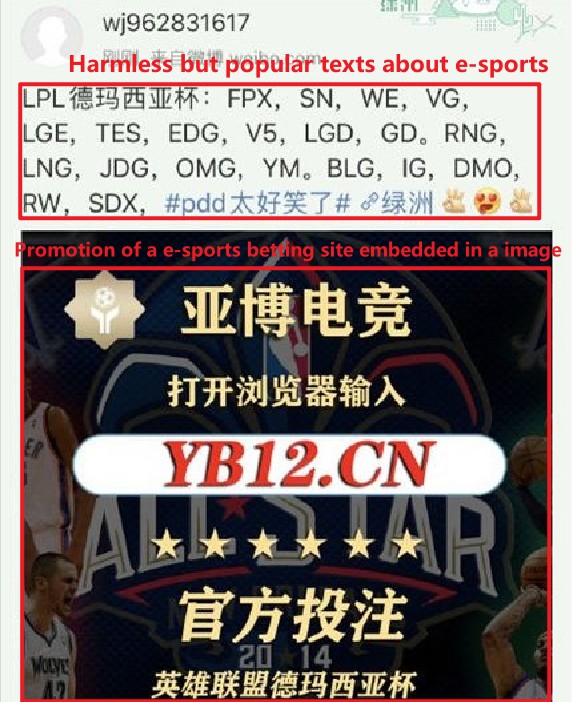}
    }
    \hfill
    \subfigure[Twitter.]{
    \label{fig:other_promotion_twitter}
    \includegraphics[width=.55\columnwidth]{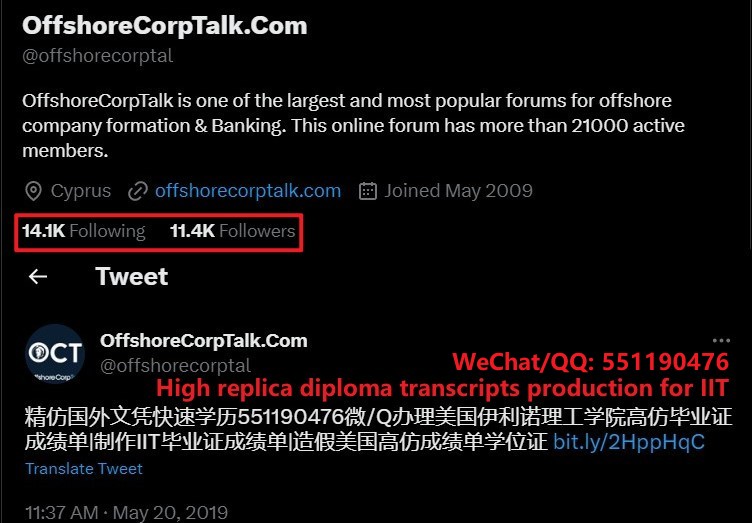}
    }
    \subfigure[Instagram.]{
    \label{fig:other_promotion_instagram}
    \includegraphics[width=.45\columnwidth]{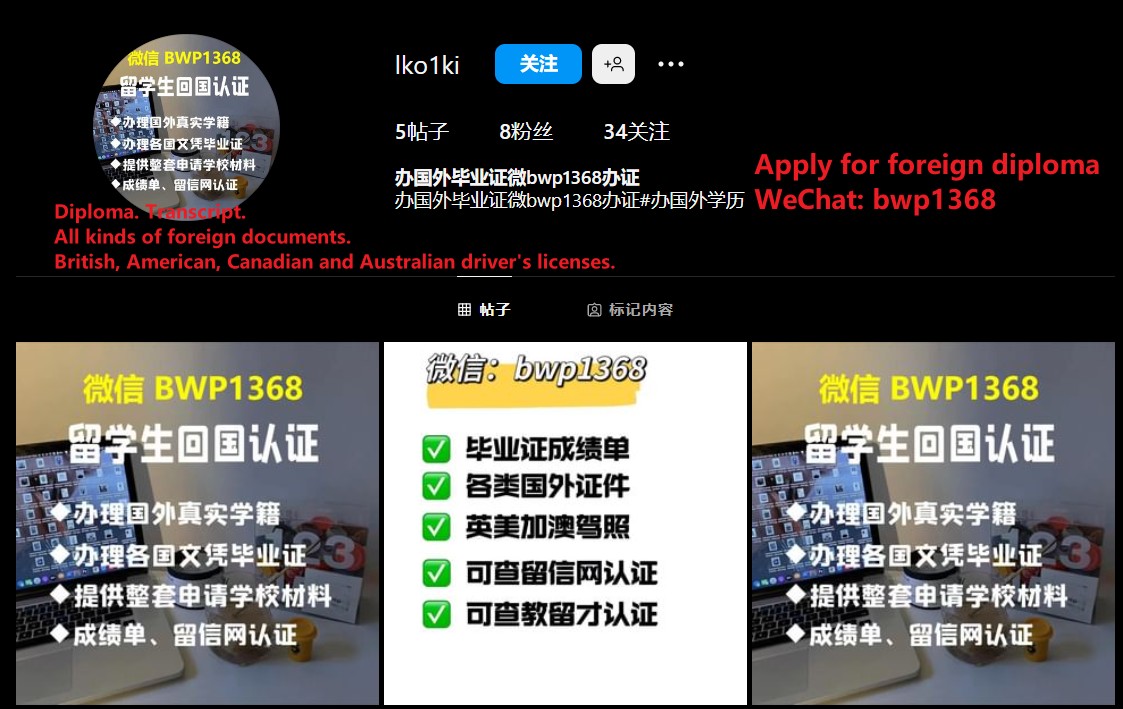}
    }
    \hfill
    \subfigure[Forum.]{
    \label{fig:other_promotion_forum}
    \includegraphics[width=.45\columnwidth]{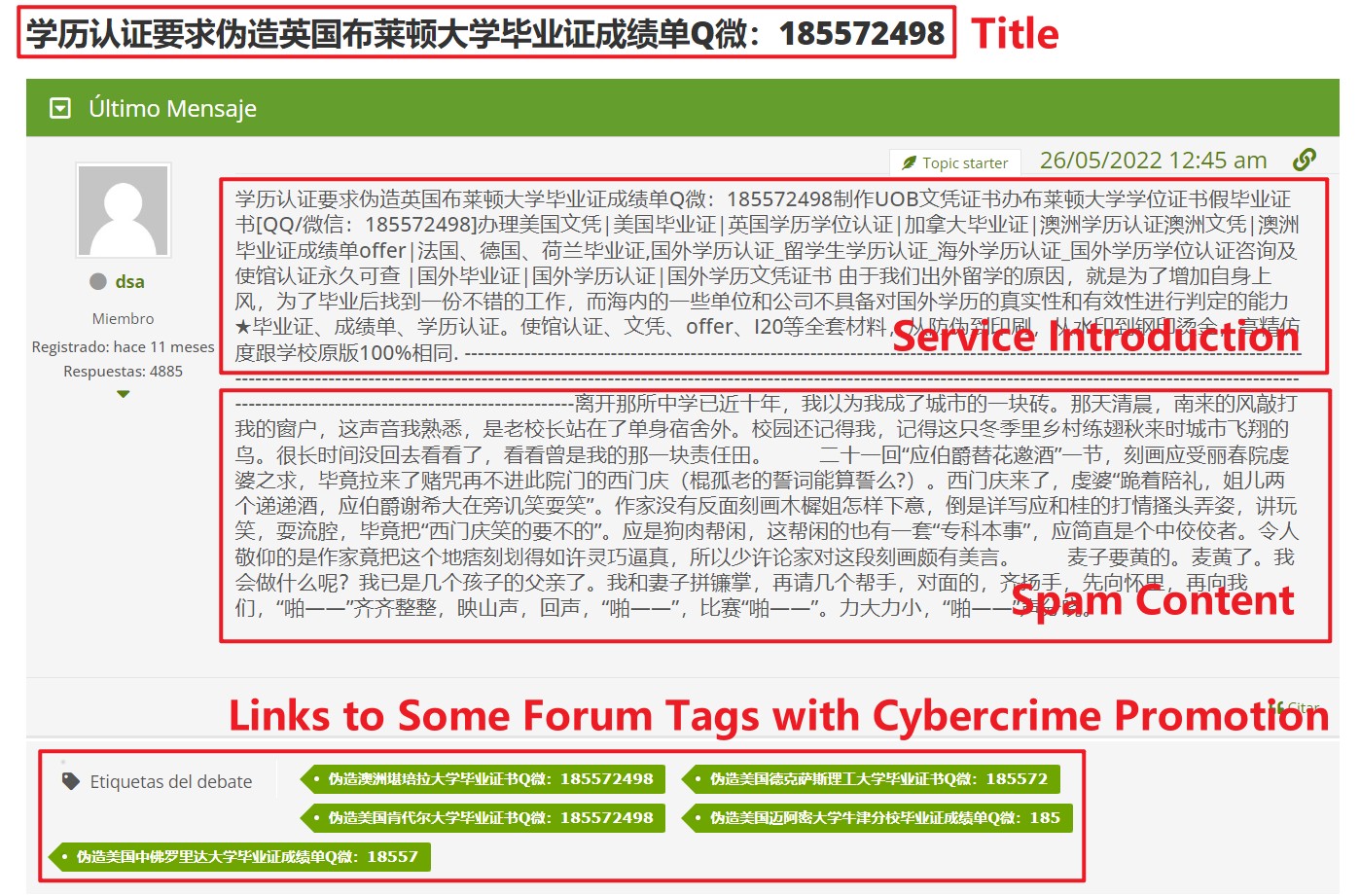}
    }
    \caption{The distribution of IPTs through other channels.}
    \label{fig:other_promotion}
\end{figure}

\subsection{IPT Distribution Through Channels Besides Search Engines}
\label{appendix:other_channel}
We also observed that miscreants distribute IPTs not only through poisoning search engines but also other popular online services, especially online social networks (OSNs) and online forums. We manually searched for the top 100 IPT contacts on various platforms. Results specific to each poisoned platform are summarized as below.

\subsubject{Weibo.}
No relevant promotions were found among the 100 contacts. However, some underground industries circumvent censorship by embedding promotions in images accompanying seemingly innocuous text (e.g. Fig. \ref{fig:other_promotion_weibo} shows a published post with gambling website promotion embedded in the image). 

\subsubject{Twitter.}
A search for these 100 contacts on Twitter resulted in 32 search results with tweets promoting IPTs. Some accounts are promotional bots, while others are compromised accounts. For instance, the account shown in Fig. \ref{fig:other_promotion_twitter} is an official forum account with 11.4K followers that is still active and shares normal forum content on a daily basis, but also contains tweets promoting IPTs.

\subsubject{Instagram.}
Only 2 illicit promotions were found due to Instagram's search limitations, which only support searching for accounts and hashtag content. However, we identified some accounts that were solely dedicated to the distribution of IPTs. These accounts prominently display their contact information in account profiles, as shown in Fig. \ref{fig:other_promotion_instagram}. 

\subsubject{Forum.}
Various forums, blogs, and Q\&A sites are exploited for the distribution of IPTs. For example, when searching Reddit for those 100 contacts, we found that 14 of them have been promoted in various posts. A representative example of such a post on a forum is shown in Fig. \ref{fig:other_promotion_forum}.

\end{CJK*}

\end{document}